\documentclass[preprint2]{aastex}

 \newcommand{\etal}{$et \; al.$~}

\def\lesssim{\mathrel{\hbox{\rlap{\hbox{%
 \lower4pt\hbox{$\sim$}}}\hbox{$<$}}}}
\def\gtrsim{\mathrel{\hbox{\rlap{\hbox{%
 \lower4pt\hbox{$\sim$}}}\hbox{$>$}}}}

\def\arcmin{\hbox{$^\prime \,$}}
\def\arcsec{\hbox{$^{\prime\prime \, }$}}

\received{2005 August 23}
\begin{document}

\title{The DEEP Groth Strip Survey VI. Spectroscopic, Variability, and
X-ray Detection of AGN}
\author{Vicki L. Sarajedini}
\affil{University of Florida, Department of Astronomy, Gainesville, FL 32611; vicki@astro.ufl.edu}
\author{David C. Koo\altaffilmark{1}, Andrew C. Phillips\altaffilmark{1}, Henry A. Kobulnicky\altaffilmark{2},
Karl Gebhardt\altaffilmark{3}, Christopher N. A. Willmer\altaffilmark{1}, Nicole P. Vogt\altaffilmark{4},
Elise Laird\altaffilmark{5}, Myungshin Im\altaffilmark{6}, Sarah Iverson\altaffilmark{7}, and 
Wanessa Mattos\altaffilmark{8}}
\altaffiltext{1}{UCO/Lick Observatory, University of California, Santa Cruz, CA 95064}
\altaffiltext{2}{University of Wyoming, Department of Physics and Astronomy, Laramie, WY 82071}
\altaffiltext{3}{University of Texas, Department of Astronomy, Austin, TX 78723}
\altaffiltext{4}{New Mexico State University, Department of Astronomy, Las Cruces, NM 88003}
\altaffiltext{5}{Astrophysics Group, Imperial College London, London, SW7 2AZ, UK}
\altaffiltext{6}{Astronomy Program, SEES, Seoul National University, Seoul, Korea}
\altaffiltext{7}{University of Iowa, Department of Physics and Astronomy, Iowa City, IA 52242}
\altaffiltext{8}{University of Florida, Gainesville, FL 32611}

\begin{abstract}

We identify active galactic nuclei (AGN) in the Groth-Westphal Survey Strip (GSS)
using the independent and complementary selection techniques of optical spectroscopy
and photometric variability.  We discuss the X-ray properties of these AGN using
Chandra/XMM data for this region. 
From a sample of 576 galaxies with high quality spectra we identify 31 galaxies with AGN 
signatures.  Seven of these have broad emission lines (Type 1 AGNs).  We also identify 26 
galaxies displaying nuclear variability in HST WFPC2 images of the GSS separated by $\sim$7 years.
The primary overlap of the two selected AGN samples is the set of broad-line AGNs, of which 80\%
appear as variable.  Only a few narrow-line AGNs approach the variability threshold.
The broad-line AGNs have an average redshift of $<$z$>$$\simeq$1.1 while the other spectroscopic AGNs
have redshifts closer to the mean of the general galaxy population ($<$z$>$$\simeq$0.7).  
Eighty percent of the identified broad-line AGNs are detected in X-rays
and these are among the most luminous X-ray sources in the GSS.  Only one 
narrow-line AGN is X-ray detected.  Of the variable nuclei galaxies within the X-ray survey, 
27\% are X-ray detected.  We find that 1.9$\pm$0.6\% of GSS galaxies to V$_{gal}$$=$24 are broad-line 
AGNs, 1.4$\pm$0.5\% are narrow-line AGNs, and  4.5$\pm$1.4\% contain variable nuclei.
The fraction of spectroscopically identified BLAGNs and NLAGNs at z$\sim$1 reveals a
marginally significant increase of 1.3$\pm$0.9\% when compared to the local population.

\end{abstract}


\section{INTRODUCTION}

Active galaxies, galaxies accreting significant material onto a central supermassive
blackhole, are some of the most intriguing objects in the universe.
The first active galaxies were identified based on their extreme
luminosities, emission line properties and ability to
vary in luminosity over short time periods.
Since then, several selection techniques have been employed to
identify and study the active galaxy population. 
However, these techniques also imposed redshift limits on the
selected samples.  Quasars, the most luminous class of active
galactic nuclei (AGN), can
be identified out to high redshifts using most selection techniques, while
the intrinsically fainter AGNs (e.g. Seyfert galaxies) are mostly limited to local
samples. 

This has led to a prominent gap in our current understanding of the 
AGN phenomenon; the
connection between bright QSOs which exist in large numbers at redshifts
of z$\simeq$2--3 and low-luminosity Seyferts
which have historically been studied at much lower 
redshifts.  Understanding how the population of faint AGNs evolves with redshift
is of particular importance for determining the frequency and
total space density of AGNs at earlier epochs.  This has
obvious implications for determining their total contribution
to the X-ray, IR and UV backgrounds.  Large numbers of low-luminosity
AGNs have been proposed to explain the ionization of the intergalactic
medium at high redshift (Steidel \& Sargent 1989), although
recent observational and theoretical results differ on the amount of their
true contribution (Barger \etal 2003a; Schirber \& Bullock 2003).
Additionally, the faint
end of the AGN luminosity function (LF) is an important constraint on evolutionary models
such as pure luminosity and luminosity-dependent
density evolution (e.g. Boyle \etal 2000).

The AGN luminosity function has been well studied out to high redshifts 
for QSOs (e.g. Hartwick \& Schade 1990) and several works measure the
local luminosity functions for Seyfert galaxies (Cheng \etal 1985;
Huchra \& Burg 1992; Koehler \etal 1997;
Ulvestad \& Ho 2001; Londish \etal 2000; Maia \etal 2003; Hao \etal 2005).
Most Seyfert LFs have been based on optical spectroscopically selected 
nuclei.  However, spectroscopically selected samples of AGNs at high redshift
are harder to obtain due to the inevitably lower signal-to-noise spectra,
combined with the lack of suitable diagnostic emission lines in the
optical portion of the spectrum that allow identifying AGNs. 

Another identifying characteristic of active galaxies is their variability.
QSOs and Seyfert galaxies have long been known as variable
objects, with significant optical flux changes occurring on
timescales of months to several years.  While variability has
been primarily used to identify QSOs
(Hawkins 1986; Koo \etal 1986;  Hook \etal 1994), Seyfert 1 galaxies
have also been shown to vary in broad-band photometric surveys (e.g. Peterson
\etal 1998).
A survey for variable sources in
SA57 revealed many optically extended, Seyfert-like galaxies
(Bershady \etal 1998) which generally had higher variability
amplitudes than the more luminous QSOs.  This result suggests that
variability is a good technique for selecting
intrinsically faint QSOs and Seyfert nuclei.   

The purpose of this study is to employ two independent and
complementary selection techniques to identify Seyfert-luminosity AGNs
to redshifts of z$\sim$1. 
We analyze the Deep Extragalactic Evolutionary Probe (DEEP)
spectroscopic survey data
of the Groth-Westphal Survey Strip (GSS; Groth \etal 1994) to 
detect and measure emission lines to determine the ionizing
source within the galaxy and thus probe the population of active
galaxies to z$\sim$1.  We further select candidate AGNs based on nuclear
variability.  Three-quarters of the GSS was re-imaged with HST 
$\sim$7 years after the original WFPC2 images were obtained. 
We identify galaxies whose nuclei have significantly varied over this 
time interval in order to select potential active galaxies.

Thus, we compare/contrast the types of AGN identified
using these two established optical selection techniques.  In addition,
we investigate how the AGN identified via optical selection manifest themselves
in deep X-ray surveys of the GSS.  This work constitutes the first comparison 
of these three AGN selection techniques for z$\sim$1 galaxies and provides
an important benchmark for future studies of AGN in deep surveys.

We describe the spectroscopic observations and line measurements in \S 2.
In \S 3, we discuss the HST photometric survey and the identified variable galaxies.
The selected AGN/AGN candidates are discussed and 
compared to typical GSS galaxies in \S 4 and \S 5.  In \S 6, 
the X-ray sources in the GSS are discussed and compared to the results of the spectroscopic and
variability surveys.  We
compute the relative fraction of Type 1 and 2 AGNs in \S 7 and
summarize the results in \S 8.

\section{THE SPECTROSCOPIC SURVEY}

The Deep Extragalactic Evolutionary Probe (DEEP; Koo \etal 1996;
Simard \etal 2002; Vogt \etal 2005; Weiner \etal 2005) 
is a galaxy redshift survey
designed to study the formation and evolution of distant field galaxies
and large scale structure to z$\sim$1.
This paper uses the first phase of DEEP (DEEP1) which was carried
out using the Low-Resolution Imaging Spectrometer (LRIS; Oke \etal 1994)
on the Keck telescopes.  The second phase of DEEP (DEEP2; Davis \etal 2003) 
is currently under way and uses the DEIMOS spectrograph 
(Faber \etal 2005) on the Keck 2 telescope.
The Groth Survey Strip is one of the
fields targeted by DEEP for spectroscopic observations and
comprises 28 contiguous WFPC2 fields located
at 14h17m+52, imaged in the V(F606W) and I(F814W) filters.
The fields are numbered 4 through 31 diagonally across the sky.
The GSS object IDs used throughout this paper indicate the GSS field (first
2 digits) and the Wide Field CCD number (third digit) in which each
object was identified.

Optical spectra were obtained through multi-slit masks for 813 objects
in the GSS between 1995 and 2001.  
The selection of objects is extensively discussed by
Vogt \etal (2005), while the description of the reduction of spectroscopic
data and redshift determination are found in Weiner \etal (2005).
In summary, each mask was observed with a blue
and red grating for a combined wavelength coverage of $\sim$5000 --
8200 $\AA$ depending on the slit placement on the mask.
The spectral resolution is FWHM=2.9 $\AA$ in the blue and 4.2 $\AA$
in the red corresponding to 2 -- 3 pixel sampling per resolution element.
Typical
exposure times were one hour per grating setup with fainter targets 
exposed for a few hours.  
Since most galaxies have sizes comparable to the seeing disk
($\sim$1$\arcsec$), 
spectral information along the spatial direction is not available in most 
cases and a one-dimensional spectrum has been produced for each object
by summing several pixels along the spatial axis. 
Six-hundred and twenty galaxy spectra were obtained in the survey and
576 have high enough signal-to-noise to yield redshifts with confidence   
levels of 90\% and higher and allow measuring spectral line parameters
(fluxes, widths and equivalent widths).  For the purpose of
identifying active galaxies, we measure several emission and
absorption lines indicative of AGN activity.
We considered only emission/absorption lines with signal-to-noise measurements
greater than 5:1 in this paper.

The line measurements were performed using two independent fitting
techniques.  The primary fitting algorithm was
developed by one of the authors (A. C. P.)
and employs a maximum likelihood estimation to perform
an automated Gaussian fit to emission/absorption lines.
The extracted sky spectrum was used to determine the noise
at each position along the spectrum. 
For about 10\% of the galaxies, however, the sky spectrum
necessary for the error determination was not available in the DEEP
database.  For these objects we manually fit the line profiles
using SPLOT in IRAF\footnote{
IRAF is distributed by the National Optical Astronomy Observatories,
which are operated by the Association of Universities for Research
in Astronomy, Inc., under cooperative agreement with the National
Science Foundation.} as described in Kobulnicky \etal (2003). 
Each line was fit individually with a Gaussian
profile except for the [OII] doublet, where two Gaussian
profiles were fit simultaneously and the sum recorded.
We empirically estimate the error in the line flux as
$\sigma$$_L$=$\sqrt{12}$$\times$rms, where 12 is the number
of pixels summed in a given emission line for this resolution, and
rms is the root-mean-squared variations in an adjacent offline region
of the spectrum.  The equivalent width error is determined as
\begin{equation}
\sigma_{EW}=\sqrt{\frac{\sigma_L^{2}}{C^{2}}+\frac{L^{2}}{C^{4}}\sigma_C^{2}}
\end{equation}
where L, C, $\sigma$$_L$, and $\sigma$$_C$ are the line and continuum
levels in photons and their associated 1$\sigma$ uncertainties.
The error in the continuum is determined by fitting the baseline
regions around each line.  
We used both fitting techniques to measure a broad sample of emission and absorption lines
from several galaxy spectra in order to compare the measured values and errors.  The line
flux, width, continuum values and errors were in general agreement to within 5\%.

For equivalent width measurements, the determination of the continuum level is
particularly important.  In almost all cases, the continuum has been averaged 
over a large number of pixels in the wavelength direction, which allows this value 
to be well determined.  The typical S/N in the continuum is well above 5 around the majority
of emission/absorption lines for most sources.  Those with the weakest continuum 
measurements (just under 1 count per 1500s of exposure time) have S/N of $\sim$1.5 to 2.

\subsection{Spectroscopically Identified AGNs} 
\subsubsection{Broad-Line AGNs}

Active Galactic Nuclei can be detected in galaxy spectra through the
presence of broad, permitted lines covering a wide range of ionization.
The galaxy spectra in the 
GSS span a wide redshift range (0.05$<$z$<$3.35) and thus we
do not have complete coverage of any one particular line for the entire
set of spectra.  Therefore, we have identified and measured
several emission lines known to
be broadened in Seyfert 1 nuclei or QSOs (i.e. broad-line AGNs or BLAGNs). 
These lines include the strongest Hydrogen Balmer 
lines (H$\alpha$ and H$\beta$),
MgII ($\lambda$2800$\AA$) and CII ($\lambda$2326$\AA$).  We also
measured CIII] ($\lambda$1909$\AA$), CIV ($\lambda$1549$\AA$) and
Ly$\alpha$ ($\lambda$1215$\AA$) for the higher redshift sources. 

The definition of ``broad" for BLAGNs is not well
defined in the literature.  The criterion in terms of FWHM ranges
between 1000 and 2000 km/s (e.g. Weedman 1977, Steidel \etal 2002).
Recently, Hao \etal (2005) found a bimodal distribution of H$\alpha$ FWHMs 
in a study of over 80,000 galaxy spectra from the Sloan Digital Sky Survey (SDSS).
These results suggest that BLAGNs can be classified as
those galaxies with FWHM(H$\alpha$) $>$ 1200 km/s.  For the sake of
uniformity, we have applied this criterion to all possibly broadened emission
lines in the spectral range of the GSS galaxies to identify BLAGNs.  
In fact, the measurements discussed here show little ambiguity in the identification
of BLAGNs which display marked differences from their narrow-line counterparts.

Of the 576 GSS spectra, 563 have at least one of the necessary emission lines 
within the spectral range.  H$\alpha$ is in the
wavelength range for 129 galaxies and found in emission for 55 sources though none
displays widths greater than 400 km/s.  The H$\beta$ line is in emission
in 226 out of 410 possible galaxy spectra.  One galaxy,
283\_3452, displays broadened H$\beta$ with a velocity of 4900 km/s.
Only seven out of 201 spectra with MgII in range display this rare emission
line.  Of these seven galaxies, six are broad with velocities between
4300 and 8300 km/s.  Two of the six display considerable self-absorption of
the broadened MgII emission line preventing an accurate measurement of the
velocity width (085\_5273 and 152\_6235).  The MgII emission in 152\_6235 
is particularly weak compared to the broad emission lines in the other galaxies.
However, there is also evidence of CIII] at the blue edge of the spectrum which
also appears slightly broad.  This adds some confidence to the classification of
this galaxy as a BLAGN, though we note that this classification is less certain
than the others.  
One other galaxy with narrow MgII emission, 092\_3172, 
also displays MgII absorption lines on either side of the emission line but the overall
shape of the emission does not appear to be broad.
Two of 48 galaxies with CII in range show the
line in emission.  One of these, 083\_5407, is broadened and also
contains broad MgII.  None of the spectra contain CIV 
emission.  Only three high redshift sources contain Ly$\alpha$ in
their wavelength range.  Two of these show the line in emission with
widths less than 300 km/s.  Thus, a total of 7 galaxies, one with
broad H$\beta$, five with broad MgII alone and one with MgII and CII, display
broadened emission lines meeting our criteria.
These galaxies are listed in Table 1 in order of increasing redshift.  The spectra and images of these
objects are shown in the first 7 panels of Figures 1 and 2.

\subsubsection{[NeV] and Broadened [OIII] as AGN Diagnositics}

The presence of [NeV]($\lambda$3426$\AA$) indicates
an ionizing potential of at least 97.02 eV, too high to be
produced through star-formation alone but possible
through accretion onto a supermassive blackhole (e.g.
Hall \etal 2000).  We detect [NeV] in only 3 of the 393 galaxies
with this line within their spectral range (see Table 2
and Figures 1 and 2).
The velocity widths of the [NeV] line range from 390 to 650 km/s.
One of these, 073\_7749, shows additional Balmer absorption lines
indicative of a post-starburst galaxy.
The other two are also classified as BLAGNs based on broad H$\beta$
and MgII.
Objects 292\_3076 and 083\_0815 also show weak [NeV] emission, though it
is below our signal-to-noise ratio for a significant detection.

Broadened [OIII]($\lambda$5007$\AA$) is also
an indicator of AGN activity.  The analysis of SDSS spectra (Hao 2003; Zakamska \etal 2003)
reveals
that AGNs typically have FWHM([OIII]) $>$ 400 km/s whereas widths for normal star-forming
galaxies are much lower.  The [OIII] line is in the spectral
range of 386 galaxies, being detected in 192.
Only two (142\_4838 and 292\_3076)
reveal broadened [OIII] with velocities of 450 km/s and 820 km/s
respectively and are listed in Table 2 (spectra and images in Figures 1 and 2).  
Object 142\_4838 is also a BLAGN.

\subsubsection{Narrow-Line AGNs}

A number of criteria have been used to identify narrow-line AGNs (NLAGNs),
which can be differentiated from normal star-forming galaxies using the
emission-line ratios of the most prominent optical lines such as
[OII]($\lambda$3727$\AA$), [OIII]($\lambda$4959,5007$\AA$),[NII]($\lambda$6583$\AA$),
[SII]($\lambda$6716,6731$\AA$), H$\alpha$ and H$\beta$ (e.g. Veilleux \&
Osterbrock 1987).  For a large fraction of galaxies in the GSS, however,
many of these lines are redshifted out of the optical range.  In addition,
the GSS spectra are not flux calibrated, making it impossible to compare
flux ratios for emission lines in different regions of the spectrum.  
For these reasons, most traditional
line ratio diagnostics are not applicable to our survey data.\footnote{
Since H$\alpha$ and [NII]/[SII] are close in wavelength and thus are not
significantly
affected by the lack of flux calibration, we did investigate this line
ratio diagnostic but found no convincing AGN candidates among those
galaxies with these lines in their spectra.}

A novel emission line diagnostic to identify NLAGNs proposed by 
Rola, Terlevich \& Terlevich (1997; RTT) uses 
the equivalent widths of the [OII] and H$\beta$ and allows 
classifying galaxies to z$\simeq$0.8 without the need 
flux calibration.  Two distinct zones define the AGN region of the diagram,
at EW(H$\beta$)$<$10 and EW(OII)/EW(H$\beta$)$>$3.5.  Using a sample
of local emission line galaxies, RTT find that 87\% of
the AGNs reside in these regions while 88\% of the HII galaxies fall
in the remaining region (see their Figure 3b).  The AGN and HII samples
are comprised of integrated galaxy spectra while the LINER spectra are
extracted from the nuclear regions of the galaxies.
Using z$\leq$0.3 emission-line galaxies from the Canada-France Redshift Survey (Tresse \etal 1996),
RTT find good agreement between their technique and classical diagnostic diagrams.    
The CFRS sample also contains spatially integrated spectra as DEEP1, though, with
a much lower resolution.
Although this technique does not perfectly
separate the star-forming galaxies from AGNs, it does a fairly good job of
identifying the majority of NLAGNs in a sample of emission line galaxies.

Of the 576 galaxies in our spectroscopic survey, 318 contain 
[OII] and H$\beta$ in their spectral range, but only
115 display both lines in emission with greater than 5$\sigma$ signal-to-noise.
In two cases, the continuum at H$\beta$ was negative due
uncertainties in the sky subtraction, making it impossible to measure the galaxies' equivalent
widths.
With the removal of these sources, we plot the remaining 113
on the RTT diagram in Figure 3a with a uniform correction of 
3 $\AA$ for underlying stellar absorption (Kennicutt \etal 1992,
Tresse \etal 1996).  This correction has the
effect of moving objects toward the HII/star-forming (SF) galaxy 
region of the diagram.  Based on the RTT criterion,
39 out of 113 galaxies (35\%) fall in the AGN
regions of the diagram.  This represents only those galaxies where
the 1$\sigma$ errorbars remain within the AGN regions of the diagram.

As an additional check on the validity of the AGN criterion from RTT, we
have analyzed a local
sample of emission line galaxies from the
Kitt Peak International Spectroscopic Survey (KISS)
(Salzer \etal 2000; Gronwall \etal 2004) using the RTT equivalent width diagram.
KISS is a large-area, objective prism survey of galaxies
selected on the basis of strong H$\alpha$ emission.  
At redshifts less than z=0.1, 318 galaxies from the KISS sample have follow-up
spectroscopic observations that include the
[OII] and H$\beta$ emission lines.  Twenty-six of 
these local galaxies are classified
as AGN/LINERs using classical line ratio diagnostic
diagrams (see Gronwall \etal 2002).  
Figure 3b shows the KISS star-forming galaxies (blue asterisks) and AGN/LINERs
(red triangles).  We have added 3 $\AA$ for underlying stellar
absorption as done for the GSS galaxies.   
We find that a threshold of EW(OII)/EW(H$\beta$)$>$5 
(dashed line in Figures 3a and 3b) maximizes the 
differentiation of the AGNs and normal star-forming KISS galaxy populations.
With this new threshold, galaxies in the upper 2 quadrants 
(AGN regions) yield a
44\% probability of being an AGN while the lower-right quadrant (star-forming
galaxies) yields a 1\% chance of being an AGN.  The lower-left quadrant
is a mixture of AGNs and star-forming galaxies (13\% AGN; 87\% SF).
It is unclear why there is such disagreement between the results of
RTT and the KISS emission-line galaxy analysis using their diagram.  However, we
will use the more conservative estimate of NLAGNs based on the KISS galaxy sample.
Assuming our emission-line galaxies are like the KISS sample,
the 7 GSS galaxies which lie above the EW(OII)/EW(H$\beta$)$>$5 threshold in Figure 3b 
have a 44\% chance of being NLAGNs.  An additional 15 galaxies lie
in the lower-left quandrant containing a mix of AGNs 
and star-forming galaxies.  Table 3 lists the 7 ``moderate probability" NLAGNs followed
by the 15 lower probability NLAGN candidates.  The galaxy images are shown in Figure 2.

\subsubsection{Absorption Line Selected AGN Candidates}

A possible indicator of AGN activity suggested by Becker \etal (1997)
is the presence of strong MgII and/or FeII absorption. 
Hall \etal (2000) identified 15
absorption-line selected AGN candidates from 0.73$<$z$<$1.33 among the
CNOC2 spectra.  However, they point out that MgII absorption lines
of comparable strength have also been identified in starburst galaxies 
(e.g. Storchi-Bergmann \etal 1995) and thus such sources can only
be classified as AGN candidates.  Only one of their 15 was confirmed
as a BLAGN through IR spectroscopy.  Through visual inspection, we 
identify 14 galaxies with significant MgII and/or FeII absorption among 
the 200 GSS spectra with these absorption features in their spectral
range.  We list these galaxies in Table 4 but do not consider them
strong AGN candidates since there is no further AGN evidence
in any of the spectra.

\subsection{Summary of Spectroscopic AGNs}

Using a variety of selection techniques, we have identified
active galaxies and AGN candidates among the 576 GSS galaxies with 
high quality spectroscopic data.
We find 7 BLAGNs, three [NeV] detected AGNs (two are also BLAGNs) and
two broadened-[OIII] detected AGN (one is also a BLAGN).
Based on [OII] and H$\beta$ equivalent width measurements, we have
also tried to identify those narrow-line galaxies that may harbor AGN.
We find 7 galaxies with a moderate probability of being NLAGNs (44\%) and 15 
galaxies with a marginal chance of being NLAGNs (13\% probability) which we
hereafter identify as AGN candidates. 
This results in a total of 31 unique galaxies identified as AGN/AGN candidates.
Those with the highest probability of being AGNs are the 16 galaxies identified through the
presence of broad emission lines, the moderate probability AGNs
from the EW diagram, the [NeV] detected galaxies and the broad [OIII] galaxies.  

\section{THE VARIABILITY SURVEY}
To identify possible AGNs via variability, we have analyzed
two epochs of HST images for a portion of the GSS.  
The original GSS was observed in the
spring of 1994 and consisted of 27 WFPC2 fields (numbers 4 through
6 and 8 through 31) imaged in
V (F606W) and I (F814W) with total exposure times of 2800s
in V for each field.  An adjacent field, known as
the Westphal field (GSS field number 7), 
was observed in the V band for 24,400s.  
In the spring of 2001, 18 of the GSS fields were reobserved
in the V band with exposure times of 4200s.  An additional
3 fields, including the Westphal field, were reobserved in spring
of 2002 with exposure times of 4200s (7200s for the Westphal field).
Thus, a total of 20 of the original 27 GSS fields plus the Westphal field 
have been analyzed for variability.
For the remainder of
the paper, we analyze the Westphal field separately 
due to the significant difference in exposure time.

\subsection{Image Processing and Photometry}
The original GSS images were combined and processed using
standard techniques discussed in Simard \etal (1999; 2002).
Unlike the images obtained in 1994, the second epoch of images for 
the 21 GSS fields were obtained in
dither mode to allow for the construction of higher resolution,
``drizzled" images.  For each field, 6 exposures with sub-pixel
shifts were drizzled together to produce 0.05$\arcsec$ resolution
images for the Wide Field CCDs using the STSDAS DITHER package in IRAF.
We then block averaged these images in two-by-two pixel blocks to
produce the same resolution as the first epoch images obtained
in 1994 (0.1$\arcsec$/pixel).  

We adopted the object coordinate catalogs for the GSS
described in Vogt \etal (2005).  All galaxy CCD positions
were visually inspected to ensure centering on the nucleus of
the galaxy.  In cases where the galaxy morphology was irregular,
the position was centered on the brightest pixel near the approximate center of
the galaxy.  For each of the 21
fields reobserved in 2001/2002, we mapped the object positions
from the epoch 1 fields onto the epoch 2 fields.  Several of the
epoch 2 fields were not well aligned with the original images, having
offsets of up to $\sim$80 pixels.  We used
GEOMAP and GEOXYTRAN in IRAF to determine the offsets and apply
the transformation to the epoch 1 catalogs.  A final visual inspection 
of the epoch 2 images overlaid with object positions ensured
that identical galaxy nuclei were identified in each epoch.\footnote{  
As an additional check that misaligned apertures from the different epochs
did not effect the photometric results, we later compared the magnitude
difference from epoch 1 to epoch 2 vs. the concentration index for each
source in our survey.  The concentration index is the difference between
magnitudes measured within r=0.15$\arcsec$ and 0.2$\arcsec$ apertures. 
The effects of poor centering should be more enhanced for concentrated
galaxies than diffuse galaxies.  We found no trend in the amount of variability
vs. galaxy concentration and therefore conclude that the aperture centering
in both epochs is robust and not a significant source of error.}

Only galaxies in the Wide Field (WF) CCDs were
included in our photometric survey.  The Planetary Camera (PC) CCD images
do not extend as deep as the WF CCDs due to the different pixel scale.
Additionally, the small field-of-view (FOV) offers little contribution to
the total sky areal coverage.  The overlapping nature of
the GSS fields allows for coverage of the PC FOV with one of the WF CCDs of an
adjacent field in most cases.  A total of 6604 galaxy nuclei were identified
for the photometric survey in the 20
GSS fields and 529 in the deeper Westphal field.

Aperture photometry was performed using IRAF PHOT
on the galaxies in each field and epoch.  We chose
an aperture radius of 1.5 pixels or 0.15$\arcsec$ to maximize
light from an unresolved nuclear component and minimize light
from the underlying host galaxy.  This is the approximately 
the measured FWHM of unresolved stars in the WFPC2 images.  We used
a magnitude zero-point of 22.91 as determined for V-band (F606W)
HST images (WFPC2 SYNPHOT update, May 1997). 

\subsection{Charge Transfer Efficiency Losses And Photometric Errors}

After performing aperture
photometry, nuclear variability
is determined from the magnitude difference of measurements
for a source at both epochs.  However,
charge transfer efficiency losses which have occurred over the
seven year interval produce a systematic offset that correlates with
position on the CCD and magnitude.  This is the well known CTE effect
(Whitmore \etal 1999; Biretta \etal 2001) which causes targets farther from 
the CCD readout amplifier to appear fainter than similar targets near 
the amplifier.  Sarajedini, Gilliland \& Kasm (2003; SGK) performed a similar 
analysis on images of the Hubble Deep Field separated by five years and
discuss the role of CTE losses in more detail.  This effect 
must be empirically quantified and removed from the data in order
to identify varying nuclei with statistical accuracy.

We found that differences in the slope of the CTE relation between
the CCDs were significant and could easily be determined for
each of the three WF CCDs individually.  We combined the data for
the 20 GSS fields, separated into individual datasets for objects
identified in WF CCD \#2, WF CCD \#3 and WF CCD \#4.  The Westphal 
field was fitted separately.
We adopted a two-step approach to determine
the CTE relationship with object position and magnitude.  
First, we fit a linear surface to the X and Y object
positions and the magnitude differences using SURFIT in IRAF for
each CCD.  The slopes of the relationship with X and Y were found to
be different for each of the 3 WF CCDs.  The fits were used to correct 
the magnitude difference values for position-dependent systematic offsets.
Secondly, we fit the dependence of the magnitude difference
with nuclear magnitude. 
A 3rd order fit was required to properly model the dependence in all
3 WF CCDs.  This correction was then applied to the data to correct
for the magnitude-dependent systematic offsets.  The overall corrections 
to magnitude difference range from 0.05 to 0.25 magnitudes with an average
offset of 0.15 centering the distribution around zero.

\subsection{Variable Nuclei Detected in the GSS}

The points in Figure 4 show
nuclear magnitude vs. CTE-corrected magnitude difference for galaxies
in the 20 GSS fields (Fig. 4a) and the deeper Westphal field (Fig. 4b).
The magnitude cut-off at the faint end (V$_{nuc}$=27 for the 20
GSS fields and V$_{nuc}$=28 for the Westphal field) was identified
as the point where the galaxy number counts peak.  Faintward of these
limits, the number counts sharply decline.  
With these limits, our photometric survey consists of
4512 objects in the 20 GSS fields and 357 in the deeper Westphal field.

To identify significant variables, we determine the expected photometric error for non-variable
sources.  Ignoring obvious outliers, we binned
the data along the magnitude
axis in 0.5 mag bins with larger bins at the bright end to include
a greater number of points, and fit the distribution of magnitude differences
with a Gaussian.  Each bin was well fit with a Gaussian and the sigma width
in each bin was determined.
These values were then fit with a 7th order polynomial 
to produce a smoothly varying photometric error as a function of magnitude.
The average magnitude difference (or photometric error) does not vary with
magnitude at the bright end.  We thus determined a lower limit to the 
photometric error at this end of the distribution.
Brightward of V$_{nuc}$=23.8, the minimum 1$\sigma$ limit for the photometric
error was found to be 0.058 based on
the average value of all probable non-varying sources brighter than this
limit.
In the deeper Westphal field, the minimum 1$\sigma$ photometric error
was found to be 0.046 for galaxies brighter than V$_{nuc}$=24.3. 

The solid line in Figure 4 represents the 3.2$\sigma$ threshold.
We have chosen this criterion to identify galaxies which
have undergone a significant nuclear flux change between
the two epochs.  This threshold was found to yield the most robust
selection of variables by minimizing both
the number of statistical outliers in the survey (only 0.14\%
in a normal distribution) and the number of
objects falling just at/above the threshold.
In the 20 GSS fields, 24 variable nuclei are detected (blue triangles)
from the 4512 galaxies surveyed brighter than V$_{nuc}$=27.
One of these, 283\_1832, appears
to be a supernova in the disk of the spiral galaxy 283\_1831 
(Sharon 2003) and falls beyond the y-axis limit
of the figure at $\Delta$Mag$\simeq$4.  
In the Westphal field, 3 variable nuclei (blue triangles) are detected among the 357 galaxies
surveyed to V$_{nuc}$=28.  Approximately 6.5 of the total number of variables are expected to be  
statistical outliers in a Gaussian distribution (see discussion of errors below).
Table 5 lists the 26 variable galaxies (omitting the likely supernova)
in order of decreasing nuclear apparent magnitude
with columns as follows:
(1) DEEP ID, (2) Redshift,
(3) V$_{nuc}$ internal to r=1.5 pixel aperture,
(4) Magnitude difference between 1994 and 2001/2002, and
(5) Significance of change obtained when normalized by the
expected error.  Images of the variable galaxies are shown in
Figure 2. 

A comparable variability study for the HDF-N 
(SGK) provides
a good comparison to the current GSS variability survey.  Sixteen
galaxies were identified in the HDF-N displaying significant nuclear flux changes
over a 5 year interval.  Figure 4 of SGK, however, shows that the variability
threshold for the HDF-N is $\sim$5 times lower than that for the GSS fields.   
The significantly lower photometric errors for non-varying sources in the HDF-N 
results from the first and second epoch observations being 
obtained with the same pointing and roll angle, such
that sources were located on the same CCD pixels in each epoch.  The
GSS fields suffered from large offsets and orientation differences
between the two epochs.  This resulted in less accurate removal
of CTE loss effects for the GSS.  
In addition, the dithering pattern for the second epoch 
observations resulted in differences in the combining and processing
of the individual images.  The combination of these effects resulted
in greater photometric noise for the GSS galaxy photometry
when compared to the HDF-N.  

As discussed in SGK, there are two distinct completeness issues
that affect this type of variability survey.  The first is related
to the incomplete time sampling of the variable sources we wish to detect.
Variability surveys usually image the survey field 
several times over many years (e.g. Trevese \etal 1994;
Hawkins 2002).
Depending on the quality of the data, these surveys have shown that
virtually all
known AGNs will be found to vary if observed periodically over
several years.  Our study is limited by
the fact that we have only two epochs with which to determine variability
and therefore sample just two points on the lightcurve of a varying source.
Because of this, we will be incomplete in our census of AGNs since
some varying sources will lie at magnitudes close to their
original magnitude measured in the first epoch and would thus be
undetected in our survey.

We estimate our incompleteness due to undersampling of the lightcurve
by using variability data for QSOs obtained over several years.
Because long term variability surveys for low-luminosity AGNs
have not yet been published, we use a sample of 42 PG quasars from
Giveon \etal (1999) and randomly select points along the lightcurve
separated by 6.5 years (the average time interval between the
epoch 1 and epoch 2 GSS images).  The Giveon \etal quasar lightcurves
cover 7 years with a typical sampling interval of 40 days.
Assuming the photometric error we determined for the GSS 
(solid line in Fig. 4),
we can estimate the probability that the Giveon QSOs would be detected in 
our survey.  The average change in magnitude
between two points on the QSO lightcurve separated by 6.5 years is
$\sim$0.2 mag and is consistent with the intrinsic variablility amplitude
of 0.14 mag reported in Giveon \etal (1999). 
We quantify the completeness by simulating photometric measurements of
the Giveon QSOs assuming the photometric error of our survey at a given
magnitude.  We then find the percentage of QSOs that would be detected
above the 3.2$\sigma$ threshold as a function of magnitude.
The completeness gets lower with increasing magnitude 
since the variability threshold increases at fainter magnitudes.  The
detection rate of QSOs ranges
from $\sim$45\% at the bright end of the distribution to about
3\% at the faint end (V$_{nuc}$=27).
To find the average survey completeness, we weight the
detection rate or completeness values by the number of galaxies in 0.5 mag bins. 
This yields a weighted detection rate of 16\% for survey
galaxies down to V$_{nuc}$=27.
At a limiting magnitude of V$_{nuc}$=26, 29\% of the QSOs would be detected.
At V$_{nuc}$=24 and brighter, 45\% of the QSOs would be detected.
Thus, the incompleteness due to lightcurve sampling is a strong function
of the survey's limiting magnitude.  
Relaxing our variability threshold to 2.5$\sigma$ would increase these values by
$\sim$6--7\% but introduces many more spurious variables as
statistical outliers.  

The second incompleteness issue results from the use of a fixed
aperture to detect nuclear variability.  We use a small, fixed
aperture to minimize dilution of the AGN light from the underlying
host galaxy to be more sensitive to AGNs varying within bright hosts.
The aperture size matches the FWHM of an unresolved
point source in the HST images (0.15$\arcsec$ in diameter).
The fixed aperture will include the same amount of light from
an unresolved source regardless of its redshift.  However, the
aperture will contain a larger fraction of the underlying, resolved
galaxy for higher redshift objects than for low-z sources.  Because
of this, the dilution of the nuclear light by the galaxy increases as
a function of redshift and, consequently, as a function of nuclear magnitude.
This issue is discussed in more detail in SGK, where they determine that
variability surveys with fixed aperture photometry become less sensitive
to variable nuclei at V$_{nuc}$$\simeq$26.5 to 27.5 in HST WFPC2 images.  
What this essentially means is that regardless of the depth of the
photometry in a WFPC2 field, only nuclei brighter than $\sim$27 are 
being uniformly surveyed for variable AGNs.
The fact that the photometry in the HDF-N field extended to
V$_{nuc}$=29.0 while all of the variables were found to be brighter 
than V$_{nuc}$=27.0 is consistent with this effect.   
In the deeper Westphal field we again see that although photometry
extends to V$_{nuc}$=28.0, the variables are all brighter
than V$_{nuc}$=27.0 (Fig. 4b).  Thus, we consider only those galaxy
nuclei brighter than V$_{nuc}$=27.0 in all GSS fields (including the   
Westphal field) to be uniformly surveyed for nuclear variability. 
The total number of galaxies surveyed to V$_{nuc}$=27.0 in all GSS
fields is 4672 (4512 in the 20 GSS fields and 160 in the Westphal field).

Since the distribution of magnitude differences between epochs is well fit
with a Gaussian, we expect 0.14\% of galaxies to fall outside of the
3.2$\sigma$ threshold as statistical outliers.  
With a total of 4672 galaxies, this results in $\sim$6.5 galaxies.    
If we correct for these outliers and the expected incompleteness,
the fraction of variable nuclei in our survey to V$_{nuc}$=27.0 is 
$\sim$2.6$\pm$0.6\% ((26-6.5)/0.16/4672).  
A recent study of the Hubble Ultra Deep Field identified $\sim$1\%
of the galaxies to $V$=28 as variable over a 4 month time interval
(Cohen \etal 2006).  This lower percentage may result from the fact that
larger apertures encompassing the entire galaxy's flux were used and thus may
be less sensitive to nuclear variations. 
In addition, the shorter time baseline is likely to
be less complete in identifying AGNs than longer, multi-year surveys
(e.g. Webb \& Malkan 2000). 
 
\section{COMPARISON OF THE SPECTROSCOPIC AND VARIABILITY SURVEYS}

Both the spectroscopic and the variability surveys presented here
have identified AGN candidates.  Unfortunately, neither
survey has 100\% complete coverage of the GSS and the two surveys
only partially overlap each other.  Because the entire GSS was not
reobserved in 2001/2002,
and due to the large offsets between the epoch 1 and epoch 2 images for some
of the repeated fields, not all galaxies with DEEP spectroscopy 
are included in the variability survey.  A total of 358 galaxies are in common
between the 7133 galaxies in the variability survey and 
the 576 galaxies with high quality spectra.  Figure 5 shows the
358 galaxies with both spectroscopic and variability data plotted on
a diagram similar to Figure 4 with nuclear magnitude along the X-axis and
magnitude change on the Y-axis.  
Spectroscopically identified AGN candidates from Tables 1 through 3
are indicated on the diagram as follows: BLAGN - blue filled circles;
moderate probability NLAGN identified in the upper two quadrants of Figure 3b - red triangles;
lower probability NLAGN identified in the lower-left quadrant of Figure 3b - red open triangles;
[NeV] selected - purple asterisks; and broad [OIII] selected - green squares. 

Most notable in Figure 5 is the fact that
3 of the 4 BLAGNs which fall within the bounds of the variability survey 
(blue filled circles) are
identified as significant variables.  The BLAGNs are also the
brightest variables in our survey.  One possible NLAGN candidate
(red open triangle in Fig. 5b) appears as a variable.  
Furthermore, some of the high-probability NLAGNs 
(red triangles) are near the 3.2$\sigma$ significance level in Figure 5a.  
BLAGNs are known to exhibit variability in their continua
while NLAGNs primarily display flux variations within
particular emission lines.  Since the broad, V-band images used
in our photometric survey are more sensitive to continuum flux changes,
it is not surprising that we find greater overlap with 
the spectroscopic BLAGNs while most of the NLAGNs are
not significant variables.  

\section{OPTICAL PROPERTIES OF AGNS IN THE GSS}

Figures 6 and 
7\footnote{To simplify comparisons, we have omitted the Westphal field galaxies from Figure 7 due            
to the exposure time difference with the other GSS fields.}
show the distribution of spectroscopically and variability detected AGNs,
respectively, compared to all
GSS galaxies in redshift, absolute magnitude, apparent magnitude, morphology and color.
The magnitudes shown are integrated galaxy magnitudes based on the galaxy model fitting
from Simard \etal (1999; 2002).  For those galaxies without model fits, we 
used 0.75$\arcsec$ radius aperture magnitudes.  Since galaxies without model fits are
typically fainter and smaller, these aperture magnitudes should adequately measure the
total light from the galaxy.
In Figure 6a, galaxy redshift vs. absolute magnitude (H$_o$=70, q$_o$=0.5) is shown for all GSS
galaxies in the spectroscopic survey to z$=$2 (black points).  
High probability AGNs are indicated with red squares
and AGN candidates with blue triangles.  The 7 BLAGNs are further indicated with
larger red squares.  The AGNs cover the full range of redshifts for the GSS survey, with the
mean redshift being slightly higher for the AGNs ($<$z$>$=0.73 compared to $<$z$>$=0.66 for
the GSS galaxies).  The BLAGNs have the highest average redshift ($<$z$>$=1.1) and also
have a mean absolute magnitude more than three magnitudes brighter ($<$M$_V$$>$=-22.6) 
than the general GSS sample ($<$M$_V$$>$=-19.3).  The mean luminosity of the NLAGNs is 
$<$M$_V$$>$=-19.8.  We expect
galaxy nuclei hosting AGNs to be generally brighter than non-AGN galaxies
since the active nucleus will contribute significant luminosity to the galaxy.  It has also been
found that the host galaxies of AGNs are typically brighter than normal galaxies (Ho \etal 1997).

Figure 7a is analogous to Figure 6a for variable nuclei galaxies.  Here 
we plot all galaxies in the photometric survey for which spectroscopic data also exist
(black points) such that redshifts are known and absolute magnitudes can be calculated.  
Significant variables are
indicated with red squares.  As shown in Figure 5a,  
5 variables are also included in the spectroscopic survey.
The three most luminous variables are the BLAGNs detected in our variability survey. 
The other 2 variables have absolute magnitudes of -19.0 and -17.3, fainter
than the average GSS galaxy magnitude of $<$M$_V$$>$=-19.3.   

The thick solid lines in
Figures 6b, c, and d are the apparent magnitude, V--I color, and
bulge-to-total flux histograms (normalized) of all 
GSS galaxies in the spectroscopic survey.
Figures 7b, c, and d display these same distributions for all GSS galaxies
in the variability survey.  The hatched/cross-hatched histograms overplotted
in Figures 6b -- 6d represent the spectroscopically detected 
AGN candidates (hatched region) and higher-probability AGNs 
(cross-hatched region).  In Figures 7b -- d, the hatched histograms are
variable galaxies with $>$3.2$\sigma$ significance and the 
cross-hatched histograms represent $>$4$\sigma$ variables.
The AGN histograms have been arbitrarily multiplied by a factor of
5 in Figure 6 and 40 in Figure 7 for display purposes.

The spectroscopic AGNs have brighter apparent magnitudes 
($<$V$>$=22.0) than typical GSS galaxies ($<$V$>$=23.0).
Comparing Figures 6b and 7b, the apparent magnitude distribution of the galaxies in the
photometric survey extends $\sim$1.5 magnitudes deeper than the spectroscopic survey. 
In addition, the detected variable galaxies cover the full
range of magnitudes for the photometric survey with a mean value only slightly 
brighter ($<$V$>$=24.0)
than that of all GSS galaxies ($<$V$>$=24.5).
The situation is different in the spectroscopic survey, where the AGNs are significantly
brighter than the typical GSS galaxy.
This highlights one of the strengths of the variability selection technique.  Such
surveys can extend deeper, with considerably less telescope time, and the identified
AGN candidates are not biased towards the brightest galaxies but cover the full apparent
magnitude range of the survey. 

The colors of galaxies in the spectroscopic and variability surveys are
similar (Figures 6c and 7c).  Although the mean color of galaxies in 
the variability survey is bluer
than the spectroscopic survey ($<$V--I$>$=0.63 compared to 0.79), both distributions
peak around V--I$\sim$0.6 and have a similar range.
A KS-test shows that the color distributions of the variability and spectroscopically
selected AGNs are not significantly different (67\% significance level). 
The redder colors observed for some of the AGN/AGN candidates 
may be due to dust which has been observed to be significant in Seyfert galaxies
(Malkan, Gorjian \& Tam 1998).  
The bluest AGNs in both the spectroscopic and variability selected samples are the
BLAGNs which have a mean V--I color of 0.45. 

The bulge-to-total measurements (B/T) in Figures 6d and 7d are from Simard \etal (1999; 2002).
These histograms represent only the subset of galaxies in the spectroscopic and
variability surveys that have 2-d model information. 
In both histograms, we have removed the BLAGNs that have unresolved
morphologies
(three in the spectroscopic survey and two in the photometric survey). 
Since the galaxy fitting algorithm did not include a point source component,
the model fits for these objects are erroneous.  
While the other AGN/AGN candidates appear to have good model fits, in some cases
the bulge measurement may be enhanced by an unresolved AGN component in
the galaxy light profile.  At the current time, 3-component
(disk+bulge+point source) modeling does not exist for the GSS galaxies, so that determining
the relative flux of point sources among the selected AGN candidates is beyond the
scope of this paper.

Interestingly, the morphologies of the AGN/AGN candidates are not 
significantly different from the parent population of galaxies 
in both the spectroscopic and variability surveys.  A KS-test comparing the AGN and
normal galaxy distributions confirms that they are not significantly different.
In both surveys,
the mean B/T for AGNs is similar to normal GSS galaxies. 
However, a comparison of Figures 6d and 7d reveals
that galaxies and AGNs in the variability
survey are slightly more bulge-dominated than the spectroscopic survey galaxies
($<$B/T$>$$\simeq$0.37 vs. 0.25).
This is also evident from a visual inspection of the galaxy images in Figure 2.
The galaxies and AGNs in the spectroscopic survey may favor more disk-dominated
systems since these galaxies are likely to have the strongest emission 
lines and therefore have more easily identifiable redshifts.

Only two of the AGN/AGN candidates from the spectroscopic and variability selected samples,
292\_3076 and 172\_5049,
are among the E/S0s identified in Im \etal (2002).
Among these, 172\_5049, a lower-probability NLAGN, is a
spheroid with a particularly blue color (Im \etal 2001). 
Their analysis indicates this object to be a low-luminosity
spheroidal with color contours revealing a bluer central
region than typical E/S0s.  This evidence supports the assertion
that an AGN may reside in the center of this galaxy.   

\section{X-RAY PROPERTIES OF SPECTROSCOPIC AND VARIABLE AGNS}

Active Galactic Nuclei have long been known as powerful X-ray
sources.  Recent deep X-ray surveys of the sky have revealed that much of the 
X-ray background is dominated by AGN activity.  X-ray surveys for
AGNs are not as hindered by some of the effects from which optical surveys
suffer, such as obscuration due to dust and contamination from
bright host galaxies.
About 60\% of the GSS has been observed with the XMM-Newton
and Chandra X-ray telescopes.  
GSS fields 4 through 18 were observed with XMM-Newton (PI:
T. Miyaji) and fields 4 through 14 with Chandra ACIS-I (PI: K. Nandra).  
A summary of the XMM detected sources is given in
Miyaji \etal (2004) and the Chandra detections  
are summarized in Nandra \etal (2005). 
A total of 31 XMM and Chandra detected sources are identified within 
1.5$\arcsec$ of an optical counterpart in the DEEP catalog.  Thirteen are
detected with both Chandra and XMM, 7 with XMM only and 11 with
Chandra only.  Some sources were detected in one survey and not the other
due to 1) the slightly larger FOV of the XMM survey and 2) the deeper flux limit of the
Chandra survey.  Table 6 lists the X-ray sources in order of decreasing
full-band X-ray flux with the following
columns: (1) DEEP ID, (2) offset between the X-ray and optical position, 
(3) XMM ID from Miyaji \etal, (4) Chandra ID from Nandra \etal,
(5) full-band (0.5--10keV) X-ray flux in units of 10$^{14}$ergs/cm$^2$/s, 
(6) hardness ratio (calculated as F$_X$(2-10)/F$_X$(0.5-2)), and (7)
full-band X-ray luminosity.
For sources detected with both Chandra and XMM, the Chandra coordinates,
X-ray flux and hardness ratios are given in the table.

Ten of the 31 X-ray sources with optical counterparts have
high quality spectroscopic data in the DEEP GSS survey and four of
these are BLAGNs.  Two of the three remaining BLAGNs in Table 1 are not
covered by the X-ray surveys.  Thus 80\% of the X-ray covered
BLAGNs in the spectroscopic survey are X-ray detected.  
The one non-detected BLAGN is also the faintest optical BLAGN.  Based
on the X-ray to optical flux ratio of sources detected in
deep Chandra surveys (e.g Barger \etal 2003b), the expected X-ray
flux for an object with this optical magnitude (approximately R$\simeq$24)
ranges over 2 orders of magnitude with the faintest limit at F$_X$(0.5-2)$\simeq$10$^{-16}$
ergs/cm$^2$/s.  Since this corresponds to the soft-band detection limit of the 200ks Chandra data
for the GSS, it is possible that this source is too faint to be detected.
Of the other
X-ray sources with spectroscopic information, one is a [NeV]
detected AGN (073\_7749).  None of the 14 EW ratio detected NLAGNs
which fell within the X-ray FOV were X-ray detected.  In total,
5 of the 10 X-ray sources (50\%) in the GSS spectroscopic survey show some
evidence of AGN activity based on their spectroscopic signatures.  
If we consider all spectroscopically selected AGNs
in the X-ray FOV
there are 25 galaxies, 5 of which are X-ray detected (20\%).
Considering only the highest probability AGNs (BLAGNs, moderate 
probability NLAGNs, [NeV] detected, and broad [OIII] AGNs) results
in 11 galaxies of which 5 are X-ray detected (45\%).
The BLAGNs are the most luminous X-ray sources in the survey and
are also among the softest X-ray emitters, consistent
with the majority of type I, QSO-like AGNs.  The [NeV]
detected AGN is among the hardest X-ray sources in the GSS
and is consistent with the classification of this object as 
highly obscured, type 2 AGN.  Given its redshift
and hardness ratio, we estimate an obsuring column density of
N$_H$$=$1.5$\times$10$^{23}$.

Some of the X-ray sources which were not observed in the DEEP1 survey or
did not reveal AGN signatures in their optical
spectra were followed up with near-IR spectroscopic observations summarized in
Miyaji \etal (2004).  Four of the five targeted galaxies did show evidence
of broad H$\alpha$ emission in the near-IR (see Table 7).  Since the signal-to-noise for
these observations is much lower than the optical spectroscopic data
analyzed in this paper, these results were not considered in the above statistics.
However, including these 4 NIR observed galaxies
as spectroscopically identified AGN raises the percentage
of X-ray sources with AGN spectroscopic signatures from 50\%
to 75\% (9/12; the total with AGN signatures becomes 5+4 and the
total number of X-ray sources with either optical or NIR 
spectroscopic follow-up is then 12). 
Clearly, a more complete spectroscopic survey extending into the near-IR would 
shed considerable light on the AGN nature of faint X-ray sources. 

The galaxies detected in the variability survey
that fall within the FOV of the X-ray data are shown in
Figure 8.  Of the 31 X-ray sources with optical counterparts,
19 are in the variability survey (red triangles).
Sixteen are above the flux completeness
limits and 4 of these are variable (3 in the GSS fields and one in the Westphal field).  
Two of these 4 are also BLAGNs while the other 2 are not covered
in the spectroscopic survey.  Thus, we find that 25\% (4 out of 16)
of the X-ray sources in the variability survey are significant variables.
If we consider the number of significant variables that are also
X-ray sources, we find that 27\% (4 out of 15) of variables in the
X-ray FOV are X-ray detected.   All 4 of the variables are
soft X-ray emitters which is consistent with the earlier
assertion that the optical variables are more likely to be
type 1 AGNs.  They are also among the most luminous X-ray sources.
About half of the variables that do not show X-ray emission 
are expected to have X-ray fluxes too faint for detection, based on typical X-ray to 
optical flux ratios.
Another two, 072\_2372 and 094\_6133, are larger galaxies or clumpy objects
whose optical light is probably an overestimate of the AGN light and thus
may also be too faint to be detected in the X-ray survey.  The remaining
few may be spurious variables or could have lower X-ray to optical flux
ratios like those of transition objects between AGN and quiescent galaxies
(Barger \etal 2003b).

To check for faint X-ray emission that may be just below the sensitivity
limits of the X-ray data, we have stacked the Chandra images at the
location of spectroscopically and variability selected AGNs which do
not individually appear as X-ray detections.  
Within 8$\arcmin$ of the Chandra aimpoint, there are 11 non-X-ray detected variable
sources.    
The stacked counts, using a 1.5$\arcsec$ extraction radius, do not reveal a
statistically significant detection (1.3$\sigma$).  However,
a marginally significant detection (2.9$\sigma$)
was found when stacking the 14 non-X-ray detected spectroscopic AGN/AGN
candidates.  These galaxies all come from the
[OII]/H$\beta$ EW selected sources listed in Table 3 (2 from the moderate probability
group and 12 from the lower probability group).  
The stacks were performed with the soft-band X-ray data (0.5--2 keV), where
the Chandra sensitivity is the greatest.
There was no significant detection using the hard-band data alone
or the full-band data (0.5--10 keV).  The marginal X-ray detection of the
[OII]/H$\beta$ EW selected galaxies provides some additional supporting evidence
for the AGN nature of galaxies selected via this method.  The lack of a
detection among the stacked variable sources may be due to their likely
higher redshifts and subsequent lower X-ray fluxes as discussed above.  The variable sources
used in the stack are generally about $\sim$2 magnitudes fainter than
the [OII]/H$\beta$ EW selected galaxies.

\section{DISCUSSION}

We have selected AGN/AGN candidates using two independent selection techniques;
spectroscopic selection based primarily on emission lines, and
variability selection based on significant flux changes in galactic nuclei over
a $\sim$7 year interval.  Ideally, one would construct a luminosity function (LF) for
both AGN samples for comparison with one another as well as other similar surveys.  
However, LFs constructed from our datasets are likely to be highly uncertain for
the following reasons: 
1) we do not have accurate knowledge of the luminosity of the AGN component in our
spectroscopically or variability selected galaxies, 
2) we have incomplete/non-uniform redshift coverage for the spectroscopic detection
of AGN, and 3) the variability survey incompleteness
estimate is quite high (84--55\%) due to poor photometric 
repeatability in the GSS WFPC2 images
(\S 3.3).  In the following paragraphs, we make estimates to account for these issues 
in order to broadly place our findings in context with published AGN LFs.

In the spectroscopic survey, we have selected AGN candidates among
various subsets of galaxies in which particular emission line features
could be measured.  
To estimate the total AGN fraction among GSS galaxies,
we first define the sets of probable Type 1 and Type 2 AGNs.
The Type 1 AGNs are easily defined as those displaying broad emission lines 
(Table 1).  Two of the three [NeV] detected sources are also BLAGNs.
The other galaxy spectrum does not show indications of any broad emission lines even though MgII
is within the spectroscopic range.  Therefore, we will consider the other [NeV]
detected source as a NLAGN.   One of the two broad [OIII] galaxies is also a BLAGN with
broad MgII.  The other, 
292\_3076, does not show any broad lines although H$\beta$ is within the spectral
range and appears as an absorption feature.  We will consider this galaxy also as a
NLAGN, although its exact nature is uncertain.  Thus we include one
[NeV] galaxy and one broadened-[OIII] galaxy in the NLAGN group along
with those galaxies identified via [OII]/H$\beta$ EW in Figure 3a.  

We compare the spectroscopic and variability survey galaxy samples with the general 
GSS catalog presented
in Vogt \etal (2005) containing 11,547 objects to identify any selection biases among our
samples and determine a characteristic limiting magnitude for AGN identification using both
of our selection techniques.  For comparison with Vogt \etal, who determined the survey
completeness as a function of combined V$_{606}$ and I$_{814}$ magnitudes in r=0.75$\arcsec$ apertures, 
we define the magnitude $GAL$ = (V$_{606}$+I$_{814}$)/2.
We investigate galaxy selection as a function of magnitude, 
color ((V-I) using r=0.5$\arcsec$ apertures) and light profile concentration (based on the difference
between r=0.15$\arcsec$ and 0.2$\arcsec$ aperture photometry).
The variability survey is 60 to 70\% complete over the full range of 
galaxy colors to $GAL$ = 24.  The incompleteness is largely due to the fact that only 2/3rds of
the GSS had second epoch data for the variability analysis.
Between magnitudes of 24 and 25, the completeness drops to about 50\% and becomes
a function of concentration, with low-concentration sources (i.e. low-surface brightness sources)
being less complete (34\%) than high-concentration sources (60\%).
The spectroscopic survey is generally less complete than the variability survey
and the completeness drops as a function of magnitude.  Much of the incompleteness is due to
the limited areal covereage of the DEEP1 spectroscopic survey.  At magnitudes greater than 21,
the survey is 42\% complete and drops to 30\% at 23rd magnitude.  Between 23rd and 24th magnitude
the spectroscopic survey includes just 13\% of galaxies. 
The completeness is roughly constant as a function of color but is
systematically lower for high-concentration sources
(stellar-like morphologies).  Therefore, we include a correction for the possibility of missed 
BLAGN due to this morphological bias (described below).

The total number of galaxies
in which NLAGNs could be identified includes all spectra with either [OII] and H$\beta$, [OIII],
or [NeV] lines in the spectral range, amounting to 526 of the available 576 galaxies with 
spectroscopic information.  
Using the [OII]/H$\beta$ EW diagram probabilities detailed in \S 2.1.3, and including
the [NeV] and [OIII] selected galaxies that do not also contain broad-lines, we estimate the
total percentage of NLAGNs to be 1.4$\pm$0.5\% at $GAL$$\leq$24, 2.3$\pm$0.9\% at $GAL$$\leq$23
and 2.5$\pm$1.4\% at $GAL$$\leq$22. 
The NLAGNs lie in the redshift range 0.14$<$z$<$0.87 with
an average redshift of z$\simeq$0.5.  The NLAGNs have integrated galaxy absolute magnitudes between
-22 to -18 (rest-frame B-band determined using H$_o$=70, q$_o$=0.5, and k-corrections 
assuming a power-law spectrum with $\alpha$=-1.0).

The fraction of BLAGNs in the spectroscopic survey can be estimated by first determining
the number of galaxies in the survey in which broad emission lines could be observed to a
given limiting magnitude.  To better estimate the total number of BLAGNs, we 
correct for the fact that the spectroscopic survey is less complete for point-like sources.
By determining the completeness of stellar-like objects vs. non-stellar objects in unity magnitude
bins, and multiplying by the fraction of BLAGNs identified among stellar-like objects
at each magnitude, we estimate that $\sim$3 BLAGNs are missed to $GAL$$\leq$24. 
Taking this into account, the percentage of BLAGNs among GSS galaxies is 1.9$\pm$0.6\% at $GAL$$\leq$24, 
2.5$\pm$0.9\% at $GAL$$\leq$23 and 5$\pm$1.9\% at $GAL$$\leq$22.  The BLAGNs in our
survey lie in the redshift range 0.65$<$z$<$1.3 and have integrated galaxy
magnitudes of M$_B$$\simeq$-24 to -20.

In the variability survey, the completeness corrected fraction of galaxies containing 
significant variable nuclei was found to be 2.6\% to V$_{nuc}$=27.0.  This nuclear
magnitude limit translates roughly to $GAL$ = 26.  In order to compare with the
spectroscopic survey, we compute the fraction of galaxies containing variable
nuclei to the same galaxy limiting magnitudes. 
A galaxy magnitude limit of $GAL$$\leq$24 imposes a V$_{nuc}$ limit of $\sim$25.
Recall that the completeness in detecting nuclear variability changes as a function
of magnitude.  At V$_{nuc}$$\leq$25, the weighted completeness is 50\%.  Correcting
for this incompleteness and statistical outliers, the percentage of galaxies containing
nuclear variables is 4.5$\pm$1.4\% .  Following the same procedure, we find the fraction
of variables galaxies to be 5.4$\pm$2.4\% for $GAL$$\leq$23 and 8.3$\pm$4.9\% for 
$GAL$$\leq$22.
The fraction of variables in GSS galaxies is systematically greater than the fraction
of BLAGNs by a few percent, though the small numbers and subsequent large errorbars on
these percentages do not make them significantly so.  If the variables are primarily 
Type 1 AGNs, as suggested by
the overlap in their samples at the bright end of the distribution, this difference is likely due to   
the ability of the variability technique to probe lower AGN/host galaxy flux ratios.  Thus, about
half of the detected variables may contain AGNs too faint to be identified on the basis
of emission lines alone.  This may be due to the intrinsic faintness of the AGN or significant
obscuration.  We do not have redshift information for the majority of variability detected AGNs.
For those with redshifts, the magnitudes lie between M$_B$$\simeq$-24 to -17.  Making
a crude estimate of the redshifts for the remaining variables (15 of the galaxies) based on V--I 
color and magnitude, the majority of variable nuclei galaxies without spectroscopic follow-up 
fall between M$_B$$\simeq$-19 to -17.  Since we find no spectroscopic BLAGNs fainter than
M$_B$$\simeq$-20, we can conclude that indeed the variability selection technique is more
sensitive to intrinsically fainter AGNs and that this is the primary explanation for the 
difference in the numbers of detected AGN. 

Previous estimates of the fraction of Type 1 and Type 2 AGNs in the local universe
reveal varying results.  Huchra \& Burg (1992; HB92) reported 1\% of local 
galaxies in the CfA Redshift Survey as BLAGNs (Seyfert 1s) and another 1\% as NLAGNs 
(Seyfert 2s).
Maiolino \& Rieke (1995) determined that at least 5\% of RSA galaxies are AGNs
with a ratio of Seyfert 1s to Seyfert 2s of 1:4.  This higher fraction 
appears to be a result of probing fainter active nuclei within galaxies
of similar magnitudes.  
Comparing the luminosity functions of Seyfert galaxies determined by HB92 to that of
Ulvestad \& Ho (2001; UH01) helps to make this clearer.  Figure 2 of UH01
shows that their survey of Palomar Seyferts has a number density that is a factor of
5--10 higher than that of the CfA survey.  Ho \& Peng (2001) show that their
median nuclear magnitude is almost 3 magnitudes fainter than that probed by
HB92.  

Our spectroscopic survey is not sensitive to intrinsically faint AGN
within host galaxies and thus our results should be more comparable to those of
HB92, where the spectra contain light from the entire galaxy and not just the nuclear
region.  It is somewhat unexpected, however, that we find no BLAGNs fainter than M$_B$=-20.
The LF of HB92 shows a steady increase from M$_B$=-22 extending to M$_B$=-18. 
For a proper comparison with HB92, we must compare the fraction of galaxies containing
BLAGNs in the same absolute magnitude range.  From Figure 1b in HB92, we see that about
2/3rds of the Type 1 AGNs lie in the absolute magnitude range of our spectroscopic BLAGN
sample (M$_B$$\simeq$-22 to -20) which constitutes 1.0$\pm$0.3\% of the CfA redshift survey
galaxies to this absolute magnitude limit.  Thus,
we find a marginally significant increase in the fraction of
galaxies containing BLAGNs in this absolute magnitude range, 
rising from 1.0$\pm$0.3\% locally to 1.9$\pm$0.6\% at z$\simeq$1.

The NLAGNs in our survey cover roughly the same absolute magnitude range as
the HB92 Type 2 Seyferts and are therefore valid for comparison.  We
find little evidence for any increase in the number of
galaxies containing NLAGNs (1\% locally to 1.4$\pm$0.5\% at $<$z$>$$\simeq$0.5).  
The increasing fraction
to higher redshifts is more significant considering a brighter magnitude limit for our spectroscopic
survey (2.5\% are NLAGNs among galaxies brighter than 22nd magnitude).
Tresse \etal (1996) found an increase in the
number of NLAGNs to z$=$0.3 in the CFRS, which consists of integrated galaxy spectra
like the DEEP survey but with lower resolution.  Using traditional line 
ratio diagnostics, they find that $\sim$8\% of galaxies are NLAGNs. 
Both the results of Tresse \etal and the present study are
sensitive to the underlying stellar absorption from the host galaxy, 
particularly the H$\beta$ emission line measurement.  Although both studies
have made estimates to account for the H$\beta$ EW measurement, uncertainties in
these estimates may explain
why the Tresse \etal fraction of NLAGNs among galaxies at z$=$0.3 is greater than that found
in the present study at a higher average redshift.
The general result is that the fraction of NLAGNs among galaxies at intermediate
redshifts (z$=$0.3 to 0.8) shows a moderate increase when compared with
the local population.  We can quantify the total AGN increase by comparing both the
NLAGN and BLAGN samples to the total number identified in the CfA redshift survey.
They find 2$\pm$0.3\% of all surveyed galaxies are AGN.  
The total fraction of spectroscopically detected AGN in
our survey is 3.3$\pm$0.8\% which yields an increase of 1.3$\pm$0.9\%. 

\section{SUMMARY}

We have used two independent and complementary optical selection techniques, spectroscopic
and variability identification, to probe the
AGN population out to redshifs of z$\sim$1 in the Groth-Westphal Survey Strip.
AGN/AGN candidates have been identified among the 576 GSS galaxies with high
quality spectroscopic data.  Thirty-one galaxies have revealed AGN spectroscopic
signatures with 9 classified as ``high probability" AGNs, 7 as ``moderate probability" AGNs, 
and 15 as ``lower probability" AGN candidates. 
Seven of these display broad emission lines (Type 1 AGNs). 
We also identify 26 variable nuclei galaxies within the GSS which have displayed
significant flux changes ($>$3.2$\sigma$) over a 6 to 7 year time interval.
The primary overlap of the two surveys is the set of BLAGNs, of which 80\%
appear as variable.  Only a few NLAGNs approach the
3.2$\sigma$ variability significance limit, indicating that variability primarily
selects Type 1 BLAGNs.   

The BLAGNs have an average redshift of $<$z$>$$=$1.1 while the other spectroscopic and
variability selected AGNs (with spectroscopic data) have redshifts closer to the mean of the general GSS
galaxy population ($<$z$>$$=$0.7).  The mean luminosity of the spectroscopically
selected BLAGNs is more than three magnitudes brighter ($<$M$_V$$>$=-22.6) than the GSS galaxy average
($<$M$_V$$>$=-19.3).
The NLAGNs are only slightly more luminous ($<$M$_V$$>$=-19.8) than the average GSS galaxy.
The variability survey probes galaxies $\sim$1.5 magnitudes fainter than the spectroscopic
survey and the AGNs identified via variability are more
representative of the entire magnitude range of galaxies.  The identified AGNs in
both surveys have colors and morphologies similar to their parent galaxy populations.

Eighty percent of the BLAGNs are detected in Chandra/XMM X-ray surveys and these are among
the most X-ray luminous sources in the GSS.  Only one of the high probability
NLAGNs is X-ray detected: a [NeV] emission line galaxy with very hard X-ray emission.  
Of the variable nuclei galaxies within the X-ray
FOV, 27\% are X-ray detected.  This can be compared to the HDF-N (SGK) where
44\% of the variable nuclei galaxies were also X-ray sources.  The higher percentage 
in the HDF-N is likely a result of
better completeness of the variability survey
and deeper X-ray obervations of the field (2Ms compared to 200ks). 
The BLAGNs and variable nuclei are among the softest X-ray sources detected.
A summary of all possible AGNs in the GSS identified via spectroscopy,
variability or X-ray detection is given in Table 7.  Along with the GSS ID
and redshift if known, the table gives the spectroscopic AGN classification,
variability in sigma units above the detection threshold and the full-band X-ray flux from either
Chandra or XMM in units of 10$^{14}$ ergs/cm$^2$/s.  

We find that 1.4$\pm$0.5\% of galaxies to V$_{gal}$$=$24 are NLAGNs while 1.9$\pm$0.6\% are
BLAGNs.  The fraction of galaxies containing variable nuclei to the same limiting magnitude is 
4.5$\pm$1.4\%.  
While it appears that variability primarily selects Type 1 AGNs (i.e. BLAGNs), the variability
survey probes intrinsically fainter galaxies (M$_B$$\leq$-17) than does the spectroscopic survey.
This is primarily due to the ability of the high-resolution HST photometry to probe flux changes 
in the central, unresolved
regions of the survey galaxies while the spectroscopic survey relies on integrated galaxy
spectra.
The fraction of spectroscopically identified BLAGNs and NLAGNs at z$\sim$1 reveals a
marginally significant increase of 1.3$\pm$0.9\% when compared to the local population.  

The results of these surveys, together with recent X-ray surveys, are helping to uncover
the population of AGNs among galaxies to z$\simeq$1.  It is clear that no one survey
wavelength or technique is able to identify all AGNs in a given field and that a
multiwavelength, multi-technique approach is the best way to identify and
study the AGN population and its evolution with redshift.
As additional multiwavelength survey data becomes available for the GSS and Extended Groth Strip region
(e.g. deep VLA, GALEX, Spitzer), a more accurate census of the AGN population to z$\sim$1
will emerge.

\acknowledgments

This paper is based on observations with the NASA/ESA Hubble Space Telescope, obtained at the 
Space Telescope Science Institute, which is operated by the Association of Universities for 
Research in Astronomy, Inc., under NASA contract NAS5-26555 
and observations with the Keck telescope, made possible by the W. M. Keck
Foundation and NASA.
Funding for DEEP was provided by NSF grants AST 95-29098 and AST 00-71198. 
This work was also supported by the STScI grant AR-09218.04.      


\clearpage

\begin{figure}
\plotone{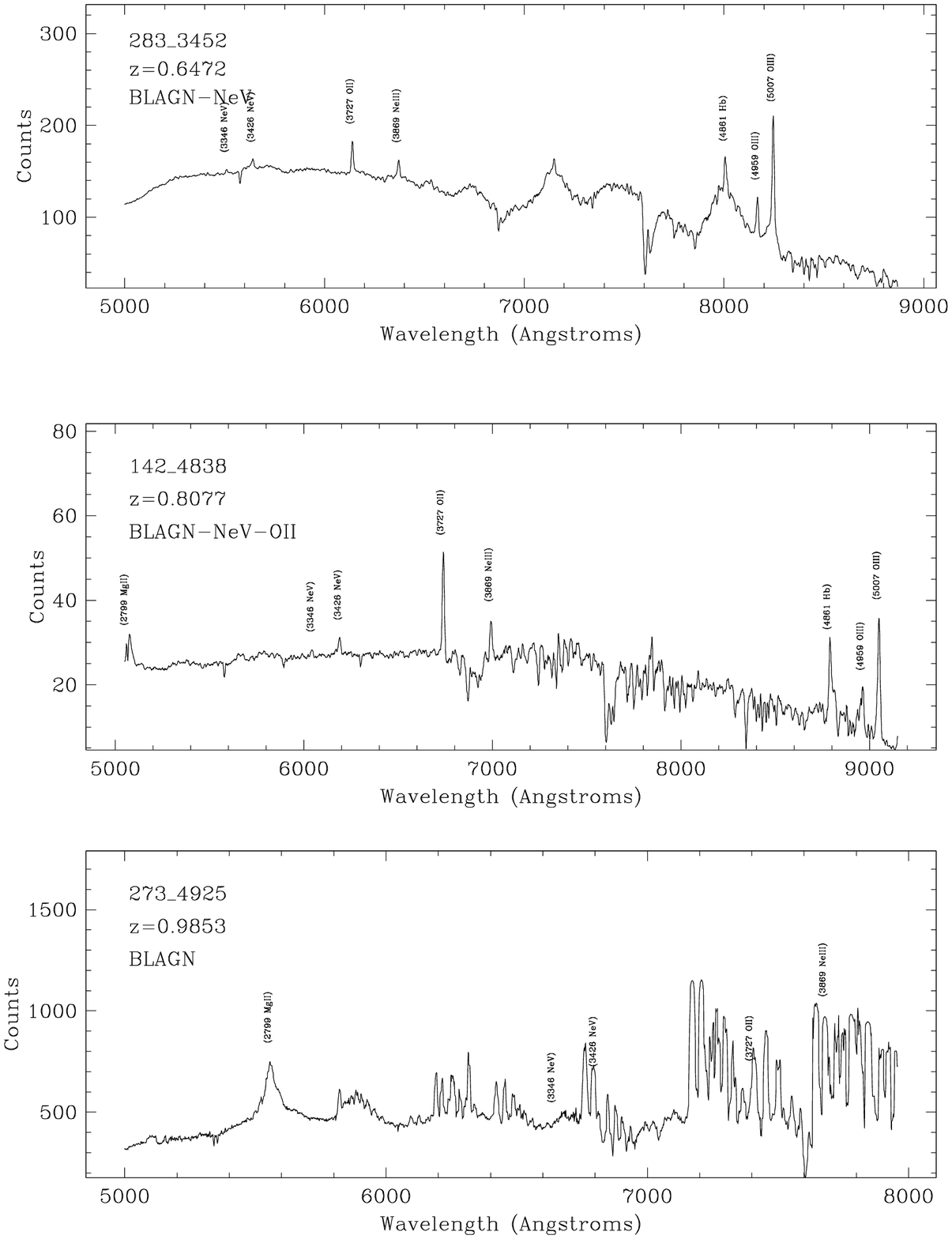}
\end{figure}

\begin{figure}
\plotone{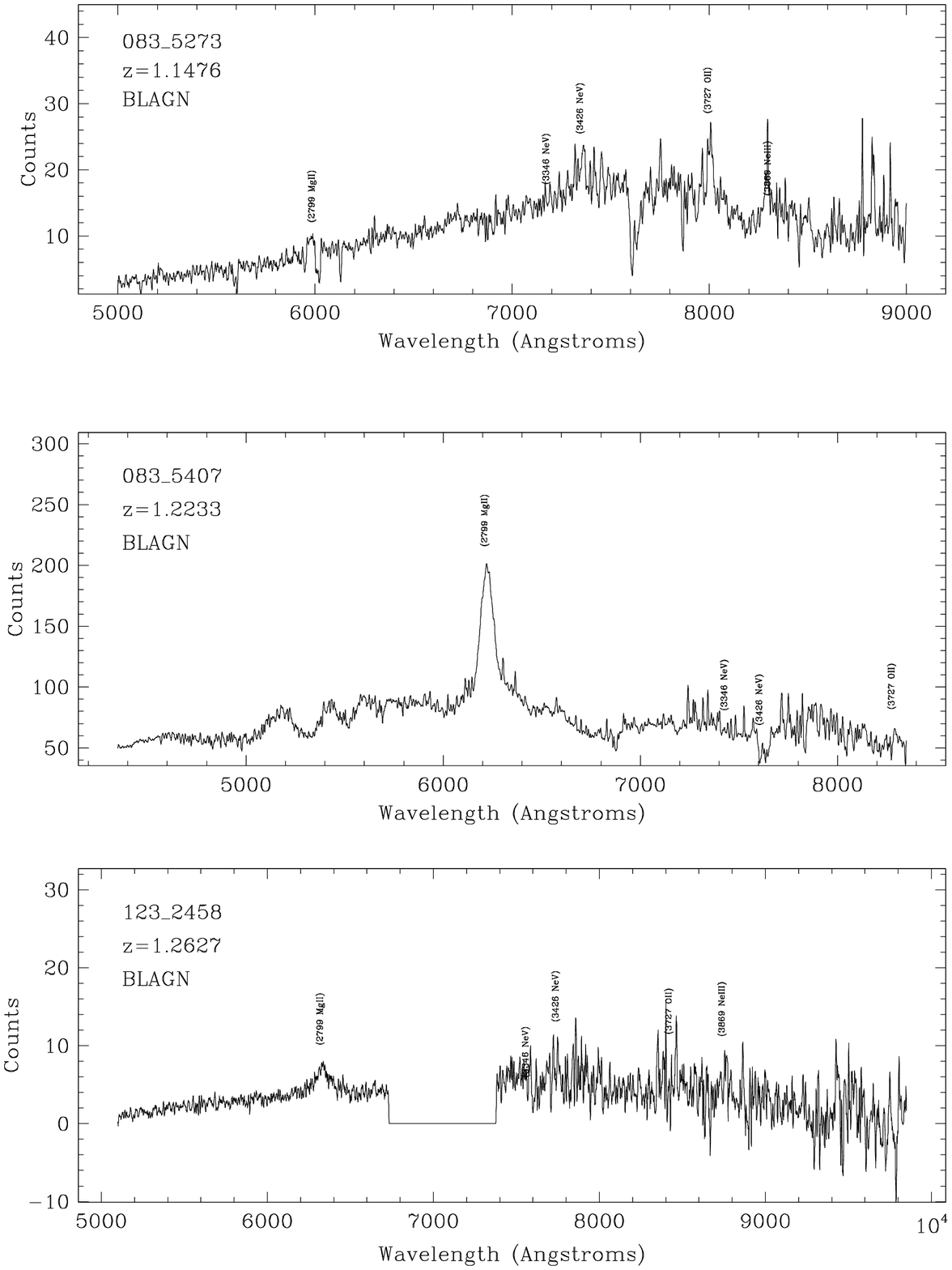}
\end{figure}

\begin{figure}
\figurenum{1}
\plotone{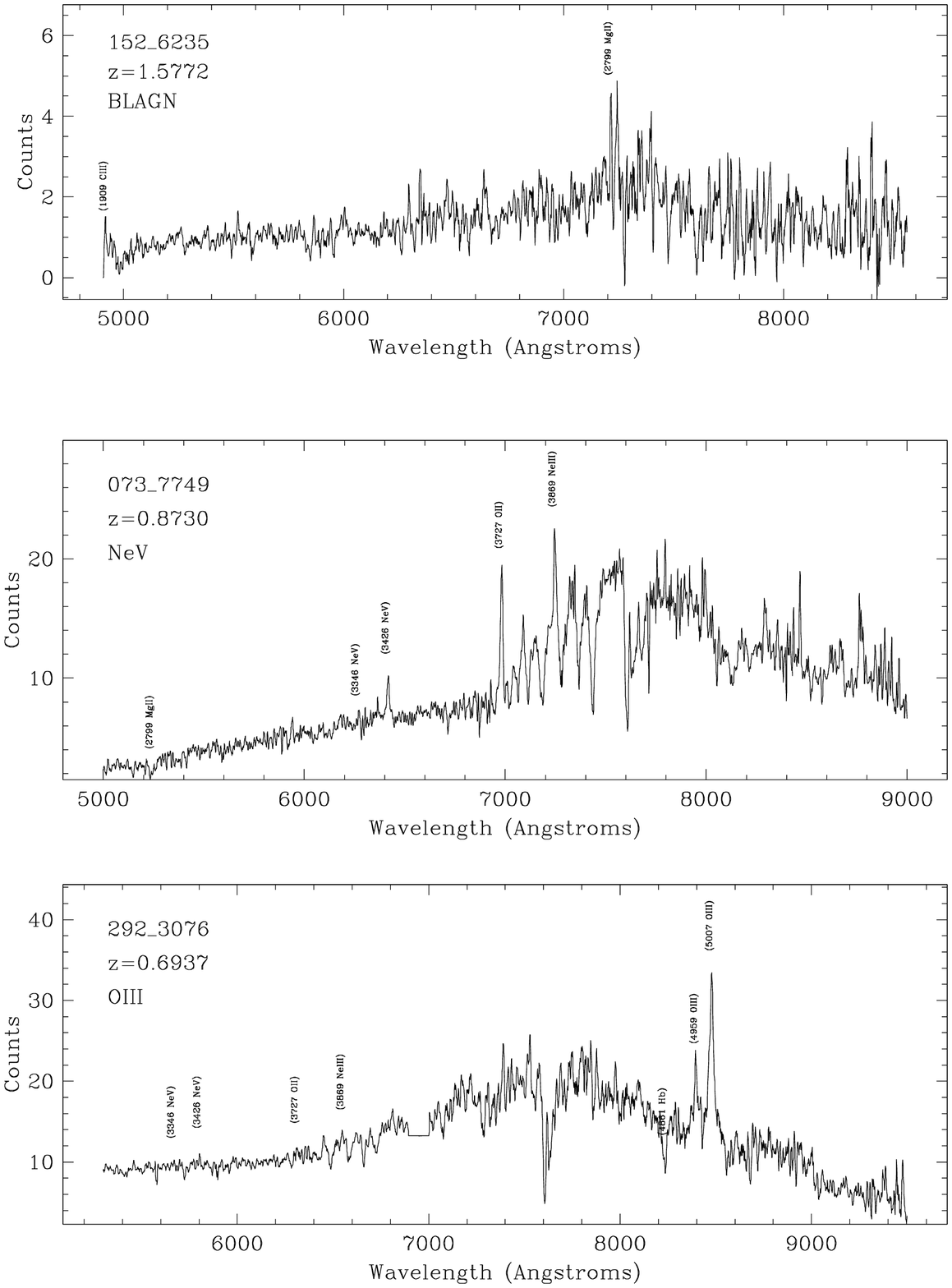}
\caption{Galaxy spectra (smoothed by 7 pixels) for AGNs listed in Tables 1 and 2.
Object 273\_4925 was observed through a set-up star hole resulting in poor
sky-subtraction.
The gap in the spectrum of 123\_2458 is a result of the non-overlapping
wavelength range of the blue and red spectra for this object.
The Y-axis is in units of ADU per pixel per 1500s exposure, where ADU=e-/2.4.}

\end{figure}


\begin{figure}
\figurenum{2}
\caption{See $www.astro.ufl.edu/vicki/DEEP1-AGN$ for Figure 2. 
V-band Galaxy images for AGN/AGN candidates listed in Tables 1, 2, 3 and 5.
Images labeled NLAGN are the 7 moderate probability NLAGNs from the upper quadrants
of Figure 3.  NLAGN-LO denotes those in the lower-left quadrant with a lower
AGN probability.  Images are 8$\arcsec$$\times$8$\arcsec$.  Additional image
scales and wavelengths can be obtained at deep.ucolick.org.}
\end{figure}

\clearpage

\begin{figure}
\figurenum{3}
\plotone{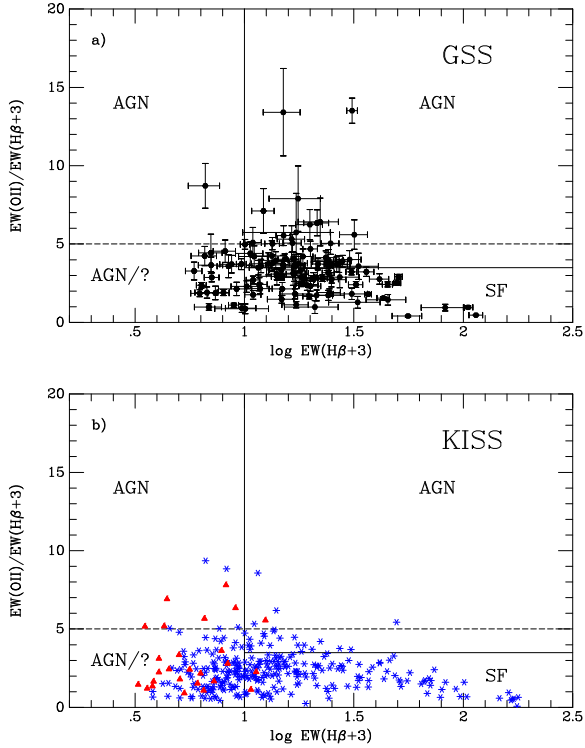}
\caption{a) GSS galaxies (113) with measureable [OII] and H$\beta$ emission.
Solid lines delineate AGNs from star-forming galaxy region according to
Rola \etal (1997). 3$\AA$ of absorption have been added to the H$\beta$ measurement.
b) Local emission-line galaxies from the KISS survey (AGNs - red triangles; star-forming galaxies
- blue asterisks) with 3$\AA$ of absorption added to the H$\beta$ measurement.
The dashed line best separates the KISS AGNs from the star-forming (SF) galaxies.}
\end{figure}

\begin{figure}
\figurenum{4}
\plotone{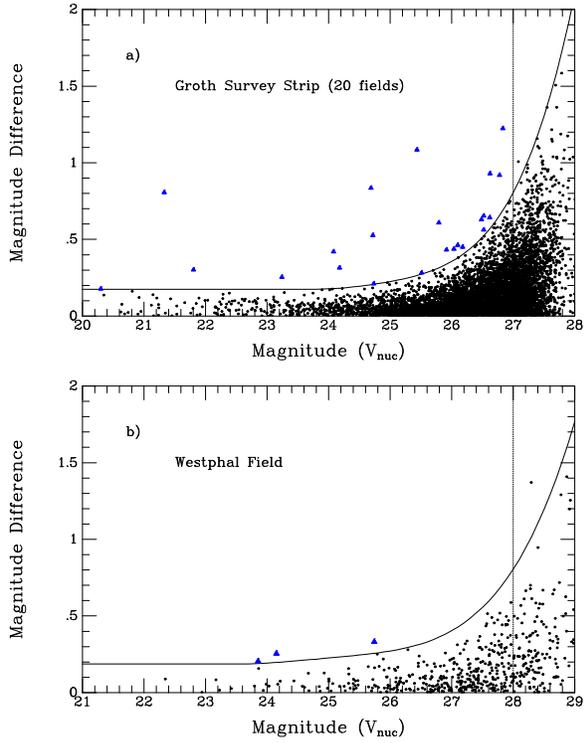}
\caption{Absolute, CTE-corrected
magnitude difference for objects in 20 Groth Strip Survey WFPC2 fields.
The solid line is the
3.2$\sigma$ limit indicating significant variables.  Objects identified as
candidate AGNs are indicated with blue triangles.
b) Same as a) for the deeper exposure time Westphal field.}
\end{figure}

\begin{figure}
\figurenum{5}
\plotone{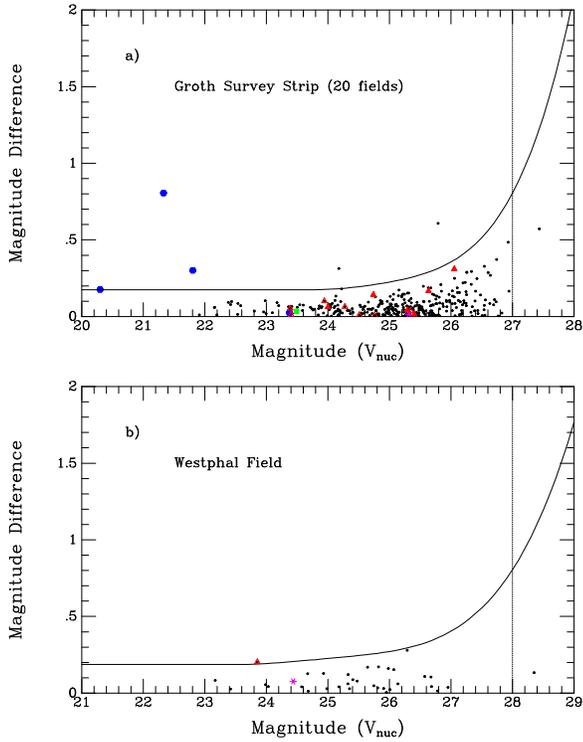}
\caption{Same as Figure 4 but for sources covered in both the
variability and spectroscopic
survey.  Spectroscopically identified AGN candidates are indicated
as follows: BLAGN - blue circles, NLAGN (RTT diagram - moderate probability AGN) - red triangles,
NLAGN (RTT diagram - lower probability AGN) - red open triangles,
[NeV] selected - purple asterisks, broadened [OIII] - green squares.}
\end{figure}

\begin{figure}
\figurenum{6}
\plotone{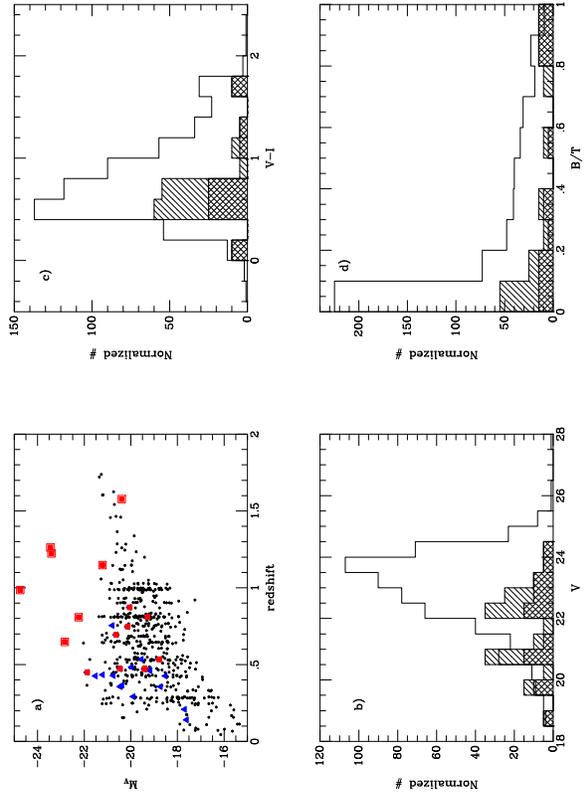}
\caption{a) Redshift vs. absolute magnitude for GSS galaxies
in the spectroscopic survey.  Red squares indicate
high and moderate probability AGNs and blue triangles are lower probability AGNs (see discussion
in text). BLAGNs are further indicated with larger red squares.  b) Histogram
of V galaxy magnitudes for all GSS galaxies (thick solid line), all selected AGN/
AGN candidates (hatched region), and high and moderate probability AGNs (cross-hatched region).
c) Histogram of V-I colors.  d) Histogram of Bulge-to-Total flux ratios.
AGN histograms are multiplied by a factor of 5 for display purposes.}
\end{figure}

\begin{figure}
\figurenum{7}
\plotone{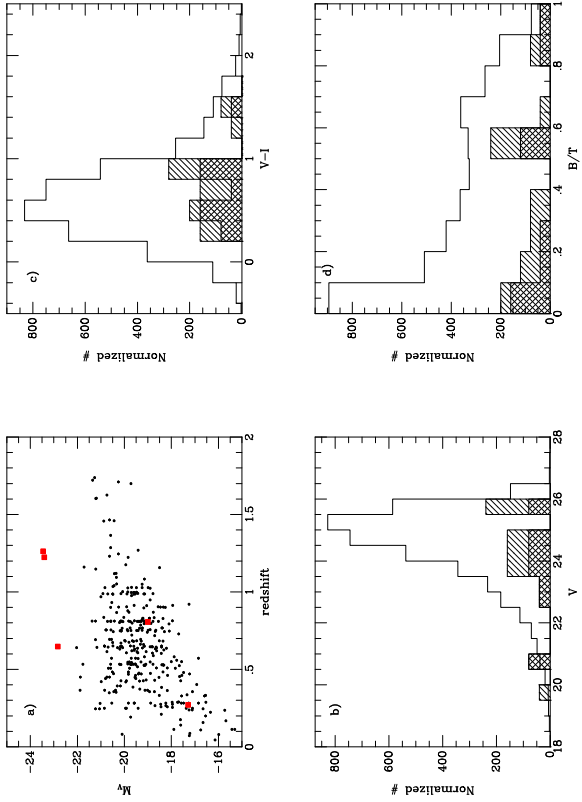}
\caption{a) Redshift vs. absolute magnitude for GSS galaxies
in the variability survey.  Red squares are
variable nuclei galaxies. b) Histogram
of V galaxy magnitudes for all GSS galaxies (thick solid line), greater than 3.2$\sigma$
variables (hatched region), and greater than 4$\sigma$ variables (cross-hatched region).
c) Histogram of V-I colors.  d) Histogram of Bulge-to-Total flux ratios.
AGN histograms are multiplied by 40 for display purposes.}
\end{figure}

\begin{figure}
\figurenum{8}
\plotone{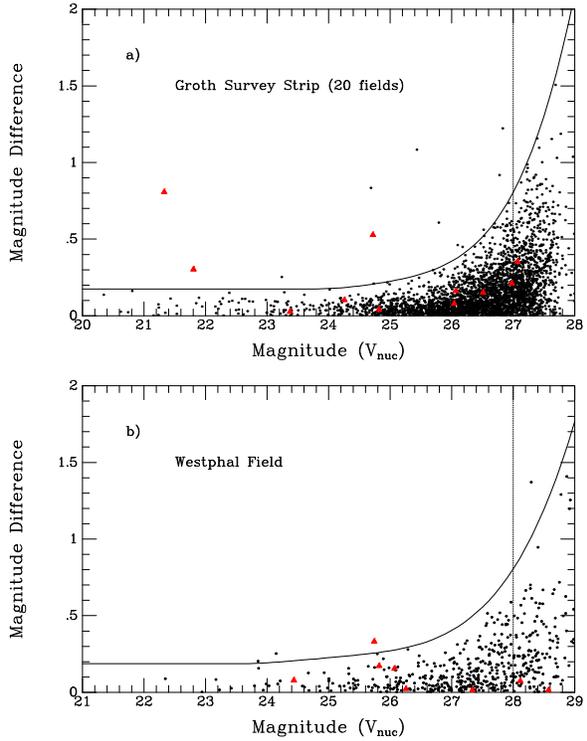}
\caption{Same as Figure 4 but for sources covered in the Chandra and XMM X-ray
surveys of the GSS.  X-ray sources are indicated with red triangles.}
\end{figure}


\clearpage

\begin{center}


\begin{deluxetable}{llccll}
\tablecaption{Broad-Line AGNs}
\setlength{\tabcolsep}{0.04in}
\tablewidth{0pt}
\tablehead{
\colhead{DEEP ID} &
\colhead{Redshift} &
\colhead{Broad Line} &
\colhead{V(km/s)} &
\colhead{m$_V$} &
\colhead{M$_B$}}

\startdata
283\_3452 & 0.6472 & H$\beta$ & 4881 & 19.63  & -22.51 \\
142\_4838 & 0.8077 & MgII & \nodata & 20.75 & -21.92 \\
273\_4925 & 0.9853 & MgII & 4322 & 18.73 & -24.41 \\
083\_5273 & 1.1476 & MgII & \nodata & 22.61  & -20.90 \\
083\_5407 & 1.2233 & MgII,CII & 8330,8373 & 20.58 & -23.08 \\ 
123\_2458 & 1.2627 & MgII & 4816 &  20.61 & -23.13 \\
152\_6235 & 1.5772 & MgII & \nodata &  24.21 &  -20.07 \\
\enddata
\end{deluxetable}


\begin{deluxetable}{llccll}
\tablecaption{[NeV] and Broadened [OIII] Selected AGNs}
\setlength{\tabcolsep}{0.04in}
\tablewidth{0pt}
\tablehead{
\colhead{DEEP ID} &
\colhead{Redshift} &
\colhead{FWHM(km/s)} &
\colhead{BLAGN?} &
\colhead{m$_V$} &
\colhead{M$_B$}}

\startdata
\cutinhead{[NeV]}
283\_3452 & 0.6459 & 390 & Yes & 19.63  & -22.51 \\
142\_4838 & 0.8065 & 540 & Yes & 20.75 & -21.92 \\
073\_7749 & 0.8730 & 650 & No & 23.13 & -19.72 \\
\cutinhead{Broadened [OIII]}
292\_3076 & 0.6937 & 820 & No & 22.00 & -20.31 \\
142\_4838 & 0.8065 & 452 & Yes & 20.75 & -21.92 \\
\enddata
\end{deluxetable}


\begin{deluxetable}{llllllll}
\tablecaption{NLAGNs Based on [OII]/H$\beta$ EW}
\setlength{\tabcolsep}{0.04in}
\tablewidth{0pt}
\tablehead{
\colhead{DEEP ID} &
\colhead{Redshift} &
\colhead{[OII] EW} &
\colhead{error} &
\colhead{H$\beta$ EW} &
\colhead{error} &
\colhead{m$_V$} &
\colhead{M$_B$}}

\startdata
284\_4709 & 0.4498 &      57.71 &    1.25 &      3.62 &    1.08 & 19.75 & -21.54 \\
193\_5426 & 0.4703 &     201.98 &   16.83 &     12.06 &    2.87 & 22.32 & -19.08 \\
223\_6264 & 0.4715 &      87.03 &   14.07 &      9.23 &    1.44 & 21.27 & -20.13 \\
154\_3015 & 0.4736 &     419.72 &    8.81 &     28.07 &    1.74 & 22.34 & -19.07 \\
133\_2106 & 0.5328 &     124.27 &    7.91 &     16.93 &    2.80 & 23.23 & -18.46 \\
084\_6809 & 0.7467 &     139.08 &    2.91 &     14.61 &    4.65 & 22.68 & -19.80 \\
084\_4515 & 0.8108 &     136.35 &    1.53 &     18.46 &    2.79 & 23.72 & -18.96 \\
\cutinhead{Lower probability NLAGNs}
083\_1919 & 0.1406 &      19.32 &    2.90 &      2.88 &    0.51 & 21.37 & -17.28 \\
102\_6577 & 0.2080 &      15.32 &    1.22 &      5.01 &    0.51 & 22.16 & -17.37 \\
072\_4040 & 0.2908 &      31.22 &    7.45 &      4.03 &    0.86 & 20.72 & -19.57 \\
172\_5049 & 0.3564 &      32.35 &    3.17 &      5.72 &    1.00 & 22.36 & -18.39 \\
173\_5210 & 0.3570 &      37.06 &    3.83 &      5.15 &    0.95 & 20.63 & -20.13 \\
213\_6640 & 0.3656 &      27.88 &    1.60 &      3.58 &    0.88 & 20.76 & -20.05 \\
072\_2372 & 0.4258 &       9.79 &    1.12 &      5.93 &    0.47 & 22.98 & -18.19 \\
092\_1962 & 0.4261 &      20.35 &    1.36 &      4.09 &    0.55 & 19.98 & -21.19 \\
113\_0815 & 0.4279 &       6.78 &    1.00 &      3.88 &    0.98 & 20.76 & -20.42 \\
094\_1054 & 0.4331 &      11.61 &    1.00 &      3.24 &    0.55 & 20.33 & -20.87 \\
103\_1520 & 0.4620 &      15.09 &    0.81 &      3.35 &    0.33 & 22.50 & -18.86 \\
103\_1811 & 0.4636 &      25.71 &    1.78 &      4.05 &    0.89 & 22.53 & -18.83 \\
072\_7471 & 0.4830 &      30.25 &    5.18 &      5.45 &    0.95 & 21.82 & -19.64 \\
093\_1519 & 0.5346 &      13.84 &    2.48 &      4.36 &    1.14 & 22.48 & -19.22 \\
084\_4521 & 0.7540 &      12.82 &    1.85 &      3.74 &    0.81 & 22.04 & -20.47 \\
\enddata
\end{deluxetable}


\begin{deluxetable}{llc}
\tablecaption{Absorption-Line Selected AGN Candidates}
\setlength{\tabcolsep}{0.04in}
\tablewidth{0pt}
\tablehead{
\colhead{DEEP ID} &
\colhead{Redshift} &
\colhead{Absorption Line}}

\startdata
153\_2622 & 0.8070 & MgII  \\
153\_2422 & 0.8072 & MgII  \\
152\_5051 & 0.8086 & MgII  \\
113\_4933 & 0.8112 & MgII  \\
093\_6661 & 0.9877 & MgII/FeII \\
104\_4809 & 0.9911 & MgII  \\
072\_2928a & 0.9972 & FeII  \\
092\_3127 & 1.0510 & MgII  \\
064\_6177 & 1.1449  & MgII \\
083\_0815 & 1.2849 & MgII  \\
212\_2577 & 1.2859 & MgII/FeII  \\
223\_7508 & 1.3658 & FeII  \\
084\_1720 & 1.6058 & MgII/FeII  \\
113\_6825 & 1.7217 & MgII/FeII  \\
\enddata
\end{deluxetable}


\begin{deluxetable}{lcccr}
\setlength{\tabcolsep}{0.04in}
\tablecaption{Variable Nuclei Galaxies in the GSS}
\tablewidth{0pt}
\tablehead{
\colhead{ID} &
\colhead{Redshift} &
\colhead{V$_{nuc}$} &
\colhead{$\Delta_{nuc}$} &
\colhead{$\sigma $}} 

\startdata
283\_3452 & 0.6472 & 20.299 &  0.176 &  3.24  \\
083\_5407 & 1.2233 & 21.330 &  0.806 & 14.83  \\
123\_2458 & 1.2627 & 21.806 &  0.301 &  5.54  \\
054\_2631 & 0.3780\tablenotemark{a} & 23.238 &  0.253 &  4.65 \\
253\_3121 & \nodata & 24.082 &  0.419 &  7.49  \\
222\_3954 &  0.2715 & 24.179 &  0.313 &  5.52  \\
052\_3372a & \nodata & 24.689 &  0.835 & 13.07  \\
053\_3424  & 1.5537\tablenotemark{a} & 24.722 &  0.525 &  8.13  \\
082\_3939  & 1.1555\tablenotemark{a} & 24.734 &  0.211 &  3.26  \\
043\_6232 & \nodata & 25.436 &  1.084 & 13.12 \\
263\_6544 & \nodata & 25.510 &  0.282 &  3.30  \\
094\_6133 & 0.8059 & 25.793 &  0.607 &  6.18  \\
254\_4642 & \nodata & 25.915 &  0.431 &  4.09  \\
284\_2213 & \nodata & 26.037 &  0.437 &  3.84  \\
124\_2138a & \nodata & 26.097 &  0.461 &  3.89  \\
124\_7332 & \nodata & 26.179 &  0.450 &  3.59  \\
274\_6339 & \nodata & 26.484 &  0.629 &  3.96  \\
214\_7315 & \nodata & 26.518 &  0.653 &  4.00  \\
124\_4831 & \nodata & 26.518 &  0.562 &  3.44  \\
222\_3930 & \nodata & 26.617 &  0.642 &  3.61  \\
272\_6943 & \nodata & 26.623 &  0.929 &  5.20  \\
172\_2078 & \nodata & 26.778 &  0.918 &  4.48  \\
042\_4373 & \nodata & 26.831 &  1.222 &  5.68  \\
\cutinhead{Westphal Field Variables}
072\_2372 & 0.4258 & 23.854 &  0.203 &  3.42  \\
074\_5765 & 0.5989\tablenotemark{a} & 24.149 &  0.253 &   4.11  \\
074\_6236 & 0.9346\tablenotemark{a} & 25.742 &  0.329 &   4.11  \\
\enddata

\tablenotetext{a}{Photometric redshift derived from multi-color
optical photometry (R. Brunner, private communications).  All
photo-z's are given in the DEEP database archive - deep.ucolick.org}

\end{deluxetable}


\begin{deluxetable}{llccccc}
\tabletypesize{\tiny}
\tablecaption{GSS X-Ray Sources with Optical Counterparts}
\setlength{\tabcolsep}{0.04in}
\tablewidth{0pt}
\tablehead{
\colhead{DEEP ID} &
\colhead{offset} &
\colhead{XMM ID} &
\colhead{Chandra ID} &
\colhead{F$_{x14}$\tablenotemark{a}} &
\colhead{Hardness Ratio\tablenotemark{b}} &
\colhead{log(L$_x$)}}

\startdata
173\_7369 & 0.50 & x11     & \nodata          & 5.2     & 3.31   & \nodata \\
083\_5273 & 0.62 & x20     & J141741.9+522823 & 5.0     & 2.92   & 44.25 \\
142\_4838 & 0.55 & x10     & J141651.2+522047 & 3.0     & 2.32   & 43.69 \\
144\_2774 & 0.34 & x9      & \nodata          & 2.8     & 0.71   & \nodata \\
142\_2752 & 1.23 & x22     & \nodata          & 2.6     & 2.28   & 43.82\tablenotemark{d} \\
083\_5407 & 0.45 & x8      & J141734.8+522810 & 2.1     & 2.06   & 43.94 \\
062\_2060 & 0.61 & x52     & J141745.9+523032 & 1.5     & 5.67   & 43.58 \\
184\_2148 & 1.29 & x55     & \nodata          & 1.5     & 3.78   & \nodata \\
164\_6109 & 1.15 & x66     & \nodata          & 1.2     & 19.17  & 43.30 \\
123\_2458 & 0.21 & x26     & J141715.0+522312 & 1.0     & 2.00   & 43.65 \\
082\_5240 & 0.33 & x44     & J141729.9+522747 & 0.92    & 2.64   & \nodata \\
074\_6236 & 1.21 & x69     & J141749.2+522811 & 0.90    & 3.59   & 43.37\tablenotemark{d} \\
053\_4446 & 0.83 & x61     & J141758.9+523138 & 0.89    & 3.67   & 42.94\tablenotemark{c} \\
184\_7960 & 1.16 & x83     & \nodata          & $<$0.86 & soft   & \nodata \\
142\_2530 & 0.46 & x40     & \nodata          & $<$0.67 & soft   & \nodata \\
074\_2638 & 1.23 & x46     & J141745.7+522801 & 0.61    & 2.59   & 42.41 \\
053\_3424 & 0.85 & x125    & J141756.8+523124 & 0.46    & 3.64   & 43.51\tablenotemark{c} \\
052\_1037 & 1.07 & x130    & J141754.2+523123 & 0.32    & 1.70   & 42.87\tablenotemark{c} \\
073\_7749 & 0.63 & x146    & J141745.4+522951 & 0.32    & 33.16  & 42.80 \\
113\_6354 & 0.03 & x133    & J141720.0+522500 & 0.31    & 4.57   & 42.58\tablenotemark{c} \\
053\_1525 & 1.01 & \nodata & J141757.4+523106 & 0.19    & 4.39   & 43.77\tablenotemark{e} \\
064\_2658 & 0.73 & \nodata & J141752.4+522853 & 0.17    & hard   & 42.60\tablenotemark{c} \\
074\_4426 & 1.27 & \nodata & J141747.0+522816 & 0.14    & 3.72   & \nodata \\
063\_6344 & 0.77 & \nodata & J141751.7+523046 & 0.11    & 2.10   & 42.46\tablenotemark{c} \\
074\_6044 & 0.62 & \nodata & J141749.2+522803 & 0.11    & 10.00  & 42.46 \\
072\_3963 & 0.64 & \nodata & J141737.3+522921 & 0.11    & hard   & 42.18\tablenotemark{c} \\
063\_4661 & 1.07 & \nodata & J141753.9+523033 & 0.11    & 3.42   & 42.46 \\
082\_3769 & 0.61 & \nodata & J141730.8+522818 & 0.099   & hard   & \nodata \\
103\_5524 & 0.42 & \nodata & J141723.6+522555 & 0.083   & 13.00  & \nodata \\
073\_2249 & 0.43 & \nodata & J141746.7+522858 & 0.076   & 2.24   & \nodata \\
072\_3223 & 0.64 & \nodata & J141739.0+522843 & 0.066   & soft   & \nodata \\
\enddata

\tablenotetext{a}{Full-band X-ray flux for Chandra sources is from Table 3, column 7
in Nandra \etal (2005) and for XMM sources is from the summation of columns 5 and 6
of Table 2 in Miyaji \etal (2004).} 

\tablenotetext{b}{A hardness ratio value of ``hard" indicates the object
was not detected in the soft (0.5--2 keV) X-ray band.  Likewise, a value of ``soft"
indicates the object was not detected in the hard (2--10 keV) X-ray band.
However, due to the greater instrument sensitivity in the soft band, a detection in this
band alone does not necessarily indicate a particularly soft source.}

\tablenotetext{c}{X-ray luminosity calculated using photometric redshift
or less certain spectroscopic redshift from DEEP database archive -
deep.ucolick.org}

\tablenotetext{d}{X-ray luminosity calculated using redshift from Miyaji \etal (2004).}

\tablenotetext{e}{Lyman break galaxy - X-ray luminosity calculated using redshift from Steidel \etal (2003).}

\end{deluxetable}


\begin{deluxetable}{lllll}
\tablecaption{Spectroscopic, Variable and X-ray detected AGNs}
\setlength{\tabcolsep}{0.04in}
\tablewidth{0pt}
\tablehead{
\colhead{DEEP ID} &
\colhead{Redshift} &
\colhead{Spectroscopic Class}  &
\colhead{Variability ($\sigma$)} &
\colhead{X-ray (F$_{x14}$)}}

\startdata
042\_4373 & \nodata & \nodata  &  5.68  & 0 \\
043\_6232 & \nodata & \nodata &  13.12 & 0 \\
052\_1037 & 0.9462\tablenotemark{a} & \nodata & \nodata & 0.32  \\
052\_3372a & \nodata & \nodata & 13.07  & 0 \\
053\_1525 & 3.0260\tablenotemark{c} & \nodata & 0.92 & 0.19 \\
053\_3424  & 1.5537\tablenotemark{a} & \nodata &  8.13 & 0.46 \\
053\_4446 & 0.6370\tablenotemark{b} & \nodata & 1.78 & 0.89  \\
054\_2631 & 0.3780\tablenotemark{a} & \nodata &  4.65 & 0 \\
062\_2060 & 0.9853 & BLAGN-IR  & \nodata & 1.5 \\
063\_4661 & 0.9976 & none & \nodata & 0.11 \\
063\_6344 & 0.8929\tablenotemark{a} & \nodata & \nodata & 0.11 \\
064\_2658 & 0.9515\tablenotemark{a} & \nodata & \nodata & 0.17  \\
072\_2372 & 0.4258 & NLAGN-LO & 3.42 & 0 \\
072\_3223 & \nodata & \nodata & 0.23 & 0.066  \\
072\_3963 & 0.7421\tablenotemark{a} & \nodata & 0.10 & 0.11 \\
072\_4040 & 0.2908 & NLAGN-LO  & \nodata & 0 \\
072\_7471 & 0.4830 & NLAGN-LO & \nodata & 0 \\
073\_2249 & \nodata & \nodata & 0.03 &  0.076 \\
073\_7749 & 0.8730 & [NeV]            & 1.20 & 0.32 \\
074\_2638 & 0.4322 & none & 2.11 & 0.61 \\
074\_4426 & \nodata & \nodata & 0.26  & 0.14 \\
074\_6044 & 0.9966 & none & 1.77 &  0.11 \\
074\_6236 & 0.9950\tablenotemark{d} & BLAGN-IR & 4.11 & 0.90  \\
082\_3769 & \nodata & \nodata & 0.85 & 0.099 \\
082\_3939  & 1.1555\tablenotemark{a} & \nodata &  3.26 & 0  \\
082\_5240 & \nodata & \nodata & 1.39 & 0.92 \\
083\_1919 & 0.1406 & NLAGN-LO & 0.22 & 0 \\
083\_5273 & 1.1476 & BLAGN          & 0.46 & 5.0 \\
083\_5407 & 1.2233 & BLAGN          & 14.83 & 2.1 \\
084\_4515 & 0.8108 & NLAGN & 0.16 & 0 \\
084\_4521 & 0.7540 & NLAGN-LO & 2.33 & 0 \\
084\_6809 & 0.7467 & NLAGN & 2.67 & 0 \\
092\_1962 & 0.4261 & NLAGN-LO & \nodata & 0 \\
093\_1519 & 0.5346 & NLAGN-LO & 0.18 & 0 \\
094\_1054 & 0.4331 & NLAGN-LO & \nodata & 0 \\
094\_6133 & 0.8059 & none &   6.18  & 0 \\
102\_6577 & 0.2080 & NLAGN-LO & \nodata & 0 \\
103\_1520 & 0.4620 & NLAGN-LO & \nodata & 0 \\
103\_1811 & 0.4636 & NLAGN-LO & \nodata & 0 \\
103\_5524 & \nodata & \nodata & \nodata & 0.083  \\
113\_0815 & 0.4279 & NLAGN-LO & 0.11 & 0\\
113\_6354 & 0.7037\tablenotemark{a} & \nodata & 0.95 & 0.31  \\
123\_2458 & 1.2627 & BLAGN          & 5.54 & 1.0 \\
124\_2138a & \nodata & \nodata &  3.89 & 0  \\
124\_4831 & \nodata & \nodata  &  3.44  & 0 \\
124\_7332 & \nodata & \nodata  &  3.59  & 0 \\
133\_2106 & 0.5328 & NLAGN & 0.50 & 0 \\
142\_2530 & \nodata & \nodata & \nodata & $<$0.67 \\
142\_2752 & 0.9830\tablenotemark{d} & BLAGN-IR & \nodata &   2.6 \\
142\_4838 & 0.8077 & BLAGN,[NeV],[OIII] & \nodata & 3.0 \\
144\_2774 & \nodata & \nodata & \nodata  &  2.8 \\
152\_6235 & 1.5772 & BLAGN          & \nodata & 0 \\
154\_3015 & 0.4736 & NLAGN & \nodata & 0 \\
164\_6109 & 0.8084 & BLAGN-IR & \nodata & 1.2 \\
172\_2078 & \nodata & \nodata  &  4.48  & 0 \\
172\_5049 & 0.3564 & NLAGN-LO & 0.98 & 0 \\
173\_5210 & 0.3570 & NLAGN-LO & 1.84 & 0 \\
173\_7369 & \nodata & \nodata &  \nodata &  5.2 \\
184\_2148 & \nodata & \nodata & 1.31 & 1.5 \\
184\_7960 & \nodata & \nodata & \nodata & $<$0.86 \\
193\_5426 & 0.4703 & NLAGN & 1.82 & 0 \\
213\_6640 & 0.3656 & NLAGN-LO & 1.11 & \nodata \\
214\_7315 & \nodata & \nodata  &  4.00  & \nodata \\
222\_3930 & \nodata & \nodata  &  3.61  & \nodata \\
222\_3954 &  0.2715 & none &  5.52  & \nodata \\
223\_6264 & 0.4715 & NLAGN & 2.16 & \nodata \\
253\_3121 & \nodata & \nodata &  7.49  & \nodata \\
254\_4642 & \nodata & \nodata  &  4.09  & \nodata \\
263\_6544 & \nodata & \nodata &   3.30  & \nodata \\
272\_6943 & \nodata & \nodata  &  5.20  & \nodata \\
273\_4925 & 0.9853 & BLAGN          & \nodata & \nodata \\
274\_6339 & \nodata & \nodata  &  3.96  & \nodata \\
283\_3452 & 0.6472 & BLAGN,[NeV]      & 3.24 & \nodata \\
284\_2213 & \nodata & \nodata &  3.84  & \nodata \\
284\_4709 & 0.4498 & NLAGN & 1.10 & \nodata \\
292\_3076 & 0.6937 & [OIII]           & 0.61 & \nodata \\
\enddata

\tablenotetext{a}{Photometric redshift}
\tablenotetext{b}{Redshift from CFRS (Lilly \etal 1995)}
\tablenotetext{c}{Redshift from Steidel \etal (2003)}
\tablenotetext{d}{Redshift from Miyaji \etal (2004)}

\tablecomments{If spectroscopic data exists but no
AGN signature was identified, the spectroscopic class is ``none".   A classification of
NLAGN is a moderate probability NLAGN and
a classification of NLAGN-LO is a lower probability NLAGN.
BLAGN-IR indicates sources showing evidence of broad H$\alpha$ reported in Miyaji \etal (2004).
If the object
fell within either the Chandra or XMM FOV but was not detected, the X-ray flux is given
as zero.}

\end{deluxetable}

\end{center}

\end{document}